\documentclass[aps,reprint,nofootinbib, amsmath,amssymb,showpacs,showkeys,superscriptaddress]{revtex4-1}

\usepackage{graphicx}  % Package for importing postscript images
\usepackage{natbib}  % Package for Bibtex
\usepackage{amsmath}  % Package for typesetting formulae
\usepackage[colorlinks=true,linkcolor=blue,citecolor=blue]{hyperref}  % Package for hyperlinking
\usepackage[top=1.5in, bottom=1.5in, left=1in, right=1in]{geometry}
\usepackage{caption}
\usepackage{amsfonts}
\usepackage{xcolor}
\usepackage{booktabs}
\usepackage{multirow}
\usepackage{bbm}
\usepackage{soul,url}
\usepackage{afterpage}

\begin{document} 

\title{Calibrating approximate Bayesian credible intervals of gravitational-wave parameters}

\author{Ruiting Mao} \affiliation{Department of Statistics,
  University of Auckland, Auckland 1010, New Zealand}

\author{Jeong Eun Lee} \affiliation{Department of Statistics,
  University of Auckland, Auckland 1010, New Zealand}

\author{Ollie Burke} \affiliation{Laboratoire des 2 Infinis - Toulouse (L2IT-IN2P3), Université de Toulouse, CNRS, UPS, F-31062 Toulouse Cedex 9, France}

\author{Alvin J. K. Chua}
\email{alvincjk@nus.edu.sg}
\affiliation{Department of Physics, National University of Singapore, Singapore 117551}
\affiliation{Department of Mathematics, National University of Singapore, Singapore 119076}

\author{Matthew C. Edwards} \affiliation{Department of Statistics,
  University of Auckland, Auckland 1010, New Zealand}
  
\author{Renate Meyer} \affiliation{Department of Statistics,
  University of Auckland, Auckland 1010, New Zealand}


\begin{abstract}

Approximations are commonly employed in realistic applications of scientific Bayesian inference, often due to convenience if not necessity. In the field of gravitational-wave (GW) data analysis, fast-to-evaluate but approximate waveform models of astrophysical GW signals are sometimes used in lieu of more accurate models to infer properties of a true GW signal buried within detector noise. In addition, a Fisher-information-based normal approximation to the posterior distribution can also be used to conduct inference in bulk, without the need for extensive numerical calculations such as Markov chain Monte Carlo (MCMC) simulations. Such approximations can generally lead to an inaccurate posterior distribution with poor statistical coverage of the true posterior. In this article, we present a novel calibration procedure that calibrates the credible sets for a family of approximate posterior distributions, to ensure coverage of the true posterior at a level specified by the analyst. Tools such as autoencoders and artificial neural networks are used within our calibration model to compress the data (for efficiency) and to perform tasks such as logistic regression. As a proof of principle, we demonstrate our formalism on the GW signal from a high-mass binary black hole merger, a promising source for the near-future space-based GW observatory LISA. 

\end{abstract}


\maketitle

\section{Introduction}\label{sec:intro}

The ground-breaking observation of a gravitational-wave (GW) signal spectacularly opened the field of gravitational-wave astronomy on the 14th of September, 2015. The ground-based gravitational-wave observatories, the Laser Interferometer Gravitational-Wave Observatory (LIGO) in Livingston and Hanford, observed a GW signal emanating from the coalescence of two stellar-mass black holes (BHs) ~\cite{abbott2016observation}. This feat was achieved through the use of sophisticated statistical signal processing algorithms \emph{and} accurate waveform templates used to filter the data stream~\cite{veitch2015parameter, abbott2020guide, abbott2016gw150914}. In a traditional (ground-based) matched-filtering search, template banks are used to detect the presence of a signal buried within the instrumental noise~\cite{kumar2014template, usman2016pycbc, ajith2014effectual, adams2016low}. Once a candidate signal in the data stream is established, stochastic sampling algorithms, such as Markov chain Monte Carlo (MCMC), are used to estimate of the parameter set that best describes the corresponding astrophysical source ~\cite{christensen2022parameter, meyer2022computational, finn1992detection}. To do this both efficiently and accurately, we require astrophysical waveform models to be both cheap to generate and sufficiently detailed to describe the fully-relativistic waveform in the data stream~\cite{Miller2005}. 

The current state-of-the-art waveform modelling technique for binary black holes with mass ratios $\lesssim 20$ is numerical relativity (NR), where the fully general Einstein equations are numerically solved for the space-time metric perturbations~\cite{lehner2014numerical, pretorius2005numerical, lehner2001numerical, boyle2019sxs}. These NR simulations are the most accurate method to date we have to generate comparable mass BHs, but can take months to generate a small number of orbits~\cite{lousto2011orbital,  blackman2015fast}. For data analysis, which relies on generating hundreds of thousands of waveforms with multiple cycles, this is computationally infeasible. In order to circumvent this, various waveform approximations have been developed that rely on a hybrid between the Post-Newtonian (PN) formalism and NR~\cite{buonanno1999effective, nagar2019efficient, cotesta2020frequency, khan2016frequency, husa2016frequency, Pratten_2020} . This has the major advantage that these waveform models are faster to generate, but with the cost that they are not truly faithful to the GW signal hidden within the data stream. Modelling errors can result in an overall reduced sensitivity in the detection of actual signals. Using non-faithful waveforms can also impact inference: we may potentially recover biased parameter estimates, and/or claim incorrectly how precise we can measure the parameters in question~\cite{Flanagan:1997kp, Miller2005, Cutler:2007mi}. Additionally, it can be shown that biases arising from approximate waveform models scale inversely in the limit of high signal-to-noise ratio (SNR)~\cite{Cutler:2007mi}. Further, waveform inaccuracies could result in systematic errors in parameter estimates and even dominate them, especially with multiple overlapping signals \cite{Hu_2023,antonelli2021noisy, pizzati2022toward, hu2023accumulating}. 

In our work, we will restrict our attention to a particular class of sources expected to be observed by the space-based gravitational-wave observatory: the Laser Interferometer Space Antenna (LISA) \cite{lisa:2017, baker2019laser}. In contrast to ground-based detectors, which are limited by seismic noise at lower frequencies, the LISA instrument will achieve optimum sensitivity in the mHz GW frequency band, providing the means to probe the rich structure of high mass binaries. One of the most promising sources of mHz gravitational radiation will be the collisions of comparable massive black hole binaries (MBHBs) with masses of $10^{5} - 10^{7}M_{\odot}$ up to redshifts of $z \lesssim 20$. Observation of these MBHBs at specific redshifts gives one a means to probe various formation channels for MBHBs ~\cite{berti2006gravitational, hughes2002untangling, sesana2005gravitational, vecchio2004lisa}. Unlike sources observed by ground-based detectors, MBHBs will be extremely loud, with SNRs up to $\sim 1000$, offering strong constraints on the parameters that govern the signal~\cite{cornish2007search, porter2008effect}. Due to the strength of these signals and the precision in which we can measure their parameters, they will provide powerful tests of general relativity (GR)~\cite{yagi2016black, barausse2020prospects, gair2013testing}.

With the immense strength of these signals, we will require exceptionally accurate waveform models to ensure that recovered parameters are not biased and that uncertainties are correctly quantified. For this reason, calibration techniques that provide a means to ``correct" inference resulting from an approximate waveform model to an exact waveform model (such as a NR simulation) may be viewed as essential. An early example of this was presented in~\cite{Flanagan:1997kp, Miller2005, Cutler:2007mi}, which showed how to estimate the bias (and thus the accurate maximum \textit{a posteriori} estimate) in the linear-signal regime. More recently, Gaussian process regression has been developed as a viable method for interpolating and marginalising over model error in the GW likelihood, thus calibrating the likelihood itself before it is used for posterior estimation \cite{Moore:2014pda,Moore:2015sza,Chua:2019wgs,Liu:2023oxw}. Similar studies can also be found with ground-based detectors, like LIGO and Virgo, as well as preparations for Cosmic Explorer and Einstein Telescope. Early Fisher studies, with an approximation to the full likelihood, may not be applicable for low SNR events. Full analyses into the full parameter space were needed when correcting the model uncertainties. Signal-specific calibration with marginalization in gravitational-wave inference was implemented~\cite{read2023waveform}. The application of the Bayesian method was also proposed to marginalize the ignorance of (unknown) higher PN order terms and give general directive calibrations~\cite{PhysRevD.108.044018}. 

When the generative model lacks computational efficiency for executing the MCMC simulation, resorting to posterior density approximation becomes an appealing option and the MCMC simulation can be avoided. This is often seen in Bayesian inference and some well-known approximations include the mean field approximation in the Variational Bayes \cite{jordan:1999} and Laplace approximation \cite{Mackay:1998, Walker:1969}. In this paper, the posterior density is normally approximated by imposing the Fisher matrix approach to the likelihood and the uniform prior. The Fisher matrix for the GW likelihood can be expressed in terms of waveform derivatives of the approximate model (which may need to be computed numerically, but using far fewer evaluations of the model than posterior sampling). It is widely used throughout GW astronomy to cheaply forecast precision statements on parameters~\cite{vallisneri2008use} and to predict biases on parameter estimates through the use of non-faithful waveform models \cite{Cutler:2007mi}. As the Fisher matrix lends itself to calculations in bulk, it is often also used to approximate and hence study a family of posteriors over a space of GW signals~\cite{Klein_2016, babak2017science, berry2019unique, colpi2019gravitational}. Thus there is plenty of motivation for methods that can calibrate the family of approximate normal posteriors obtained via the Fisher matrix.

In this paper, we introduce a novel calibration technique that approaches the problem from another angle, using the formalism presented by \cite{lee:2019}. In the Bayesian framework, the uncertainty about unknowns is probabilistically represented by the posterior distribution. When an approximate waveform generation model is used or a likelihood/posterior is approximated, the resulting inference is not exact. {\it Operational coverage} of a credible set of a parameter vector based on an approximate posterior measures how much of the exact posterior probability mass lies in this set and, it can be interpreted as an error estimate for the approximate posterior. A practical operational coverage estimator allows us to estimate an error of posterior approximation for any observed waveform generated from a prior distribution \citep{lee:2019,xing:2019}. To perform the calibration formalism, it is necessary to generate a large number of posteriors using an approximate waveform model. 
This would clearly become computationally prohibitive with expensive posterior simulation methods. Due to the usage of normal approximations via a Fisher-based formalism to the posteriors, expensive posterior simulation is not required and, the practical operational coverage estimator is acquired with a more budget-friendly computing cost.

Here, we present a practical estimator for gravitational-wave data analysis and demonstrate how to compute the calibrated credible set of an approximate posterior that corresponds to the desired exact posterior probability mass. Systematic studies usually focus on correcting a \emph{single} posterior, generated via an approximate waveform model at a specific point in parameter space~\cite{antonelli2021noisy}. Instead, we propose a method that, after a training scheme (on the prior space of samples), near-instantaneous calibrated posterior estimates can be generated over the entire signal space. In other words, we devise a scheme that can calibrate a \emph{family} of posteriors, rather than a single one.

This paper is organised as follows. In Section \ref{sec:GW_data_analysis} we set notations and introduce the data analysis concepts that will be used throughout this work. In Section \ref{sec:general_calibration_methodology}, we summarise the work of \cite{lee:2019}, outlining a framework that can be used to calibrate the statistical coverage of approximate posterior distributions to the exact posterior distribution as if parameter estimation was performed using a more faithful waveform model. In Section \ref{sec:toy_model} we demonstrate this calibration procedure using a simple toy example and finally show its general applicability on an MBHB source within the LISA framework in Section \ref{sec:massive_black_holes_application}. Our conclusions and scope for future work is presented in section \ref{sec:discussions_conclusions}.

%%%%%%%%%%%%%%%%%%%%%%%%%%%%%%%%%%%%%%%%%%%%%%%%%%%%%%%%%%%%%%%%%%%%%%%%%%%%%%%%%%%%%%%%

\section{Gravitational-wave data analysis}\label{sec:GW_data_analysis}

\subsection{Noise modelling and likelihood}

The typical time-domain data stream observed by the LISA instrument will be a combination of TDI variables $X = \{A, E, T\}$, representing the response of the LISA instrument to the plus and cross polarisations of the incoming GW source in the transverse-traceless gauge~\cite{tinto:2021, tinto:2023}:
\begin{equation}\label{eq:data_stream_X}
d_{o}^{(X)}(t) = h_{\text{e}}^{(X)}(t;\boldsymbol{\theta}_{\text{0}}) + n^{(X)}(t), \quad X = \{A, E, T\}.
\end{equation}
Here $d_{o}$ is the observed data stream, $\boldsymbol{\theta}_{0}$ are the true parameters of the true gravitational wave $h^{(X)}_{\text{e}}$, and $n^{(X)}(t)$ are noise fluctuations arising from perturbations to the LISA instrument from unresolvable GW sources and non-GW instrumental perturbations. In our work, we will perform inference on a \emph{single} waveform within the data streams $d^{(X)}$ and ignore potential multiple signals within the data stream, such as would be considered in the global fit \cite{littenberg:2023}.  We make the assumption that the noise $n^{(X)}$ in each channel is a weakly stationary Gaussian stochastic process with zero mean, coloured by the power spectral density (PSD) of their respected TDI channel. A consequence of this is that the noise $n^{(X)}$ is uncorrelated in the frequency domain, resulting in a purely diagonal noise covariance matrix $\Sigma$ ~\cite{wiener1930generalized, khintchine1934korrelationstheorie, burke2021extreme}:
\begin{align}\label{eq:noise_stream_X}
\Sigma(f,f') &= \langle \hat{n}^{(X)}(f)(\hat{n}^{(X)}(f'))^{\star} \rangle \\ &= \frac{1}{2}\delta(f - f')S^{(X)}_{n}(f')\,.\label{eq:noise_stream_X}
\end{align}
for $f\in(0,\infty)$. Here $\langle \cdot \rangle$ denotes an average ensemble over many noise realisations, $\delta$ is the Dirac delta function and $S_n^{(X)}$ is the power spectral density (PSD) of the noise process within a channel $X$. Hatted quantities refer to the Fourier transform with convention,
\begin{equation}
    \hat{h}(f) = \int_{0}^{\infty}\text{d}t \, h(t)\exp(-2\pi \text{i} f t).
\end{equation}
Assuming that the arm-lengths of the LISA interferometer are both equal and constant, it can be shown that the noise across channels $X$ is independent and thus uncorrelated \cite{tinto:2021, tinto:2023}. From equation \eqref{eq:noise_stream_X}, Whittle showed that the likelihood in the frequency domain takes the form \cite{whittle:1957}
\begin{equation}\label{eq:whittle_likelihood_noise}
    p(n) = -\frac{1}{2}\sum_{X}(n|n)^{(X)}
\end{equation}
with inner product~\cite{finn1992detection, Flanagan:1997kp}
\begin{equation}\label{eq:inner_prod}
    (a|b)^{(X)} = 4\text{Re}\int_{0}^{\infty}\text{d}f\frac{\hat{a}^{(X)}(f)(\hat{b}^{(X)}(f'))^{\star}}{S^{(X)}_{n}(f')}.
\end{equation}
Substituting equation \eqref{eq:data_stream_X} into \eqref{eq:whittle_likelihood_noise}, we obtain the usual likelihood used throughout gravitational-wave astronomy~\cite{finn1992detection, meyer2022computational, burke2021extreme}
\begin{equation}\label{eq:whittle_likelihood_signal}
p(d|\boldsymbol{\theta}) = -\frac{1}{2}\sum_{X}(d - h_\text{m}|d - h_\text{m})^{(X)},
\end{equation}
where $h_\text{m}$ are our approximate model templates, favourably quick to generate and used when inferring parameters $\boldsymbol{\theta}$. 

The signal-to-noise ratio (SNR), $\rho$, is a quantity used to determine the power of the signal when compared to noise. Within the framework of matched filtering~\cite{woodward2014probability, turin1960introduction}, the optimal matched filtering SNR takes the form~\cite{finn1992detection}
\begin{equation}\label{eq:SNR_X}
    \rho_{X}^{2} = (h_{m}|h_{m})^{(X)} = 4\int_{0}^{\infty}\text{d}f \frac{|\hat{h}(f)|^{2}}{S^{(X)}_{n}(f)},
\end{equation}
with total (squared) SNR across $X = \{A, E, T\}$ given by summing equation \eqref{eq:SNR_X} in quadrature $\rho^2 = \sum_{X}(h_{\text{m}}|h_{\text{m}})^{(X)}$. 

We now describe how we generate detector noise given that the noise is both stationary and we know, a-priori, the PSD of the noise process $n^{(X)}$.

The frequency domain equation \eqref{eq:noise_stream_X} can be discretized in the continuum limit to give the covariance of the noise between two frequency bins $f_{i}$ and $f_{j}$:
\begin{eqnarray}
\hat{\Sigma}_{ij} &=& \mathbb{E}_{d}[\hat{n}^{(X)}(f_{i})(\hat{n}^{(X)}(f_{j}))^\star]\\ 
&=& S_{n}^{(X)}(f_{i})\delta_{ij}/2\Delta f.\label{eq:discrete_noise_gen}
\end{eqnarray}
Here $f_{i} \in [0, \Delta f, \ldots ,(\frac{N}{2})\Delta f]$ is an individual frequency bin, $\Delta f = 1/N\Delta t = 1/T_{\text{obs}}$ the spacing between frequency bins, $N$ the length of the time series and $\Delta t$ the sampling interval. 

Equation \eqref{eq:discrete_noise_gen} highlights that, for stationary Gaussian noise, the frequency bins for $i\neq j$ are uncorrelated. Focusing on the diagonal elements of \eqref{eq:discrete_noise_gen}, it is possible to show that the real and imaginary components of the noise follows a Gaussian distribution:
\begin{subequations}
\begin{align}
\text{Re}(\hat{n}^{(X)}(f_{i}) &= N\left(0,\frac{S_{n}^{(X)}(f_{i})}{4\Delta f}\right) \label{eq:real_part_noise}\\
\text{Im}(\hat{n}^{(X)}(f_{i}) &= N\left(0,\frac{S_{n}^{(X)}(f_{i})}{4\Delta f}\right) \label{eq:imag_part_noise}\,.
\end{align}
\end{subequations}
To simulate noise, we thus draw components of the noise from equations \eqref{eq:real_part_noise} and \eqref{eq:imag_part_noise} given PSDs $S_{n}^{(X)}$ for each of the channels $X = \{A, E, T\}$. An exact signal is then generated, added to this specific noise realisation, and this constructs the data set $d^{(X)}(t)$ in \eqref{eq:data_stream_X}. 

As will be discussed in \ref{sec:FM}, assuming that the likelihood is consistent with the noise model, the detector noise will encode a statistical fluctuation forcing a deviation between the recovered and true parameters. In reality, the recovered parameters will \emph{not} be centred on the true parameters due to the presence of two features: waveform modelling errors and noise. The next section describes how one can compute the bias on parameters due to waveform modelling errors and statistical fluctuations due to the inclusion of noise. 

\subsection{Fisher matrix}\label{sec:FM}

In our work we will use a Fisher matrix formalism to generate approximate distributions on parameters given an observed data stream $p(\boldsymbol{\theta}|d_{o})$~\cite{finn1992detection, vallisneri2008use}. In the high SNR limit, it is expected that the distribution of parameter sets is a multivariate Gaussian defined by a mean vector and covariance matrix. Here we will show how one can approximate both the mean vector and covariance matrix using a Fisher matrix approach, rather than using costly MCMC simulations. Here we generalise the results from~\cite{Flanagan:1997kp, Miller2005, Cutler:2007mi} to account for the three LISA channels $A, E$ and T.

We denote the best fit parameter $\theta^{i}_{\text{bf}}$ as the \emph{maximum likelihood estimate} (MLE) that maximises equation \eqref{eq:whittle_likelihood_signal} 
\begin{equation}\label{eq:max_likelihood}
\sum_{X = \{A,E,T\}}(\partial_{i}h^{(X)}_{m}(t;\boldsymbol{\theta}_{\text{bf}})|d^{(X)} - h^{(X)}_{m}(t;\boldsymbol{\theta}_{\text{bf}})) = 0\,.
\end{equation}
Here $\partial_{i} := \partial/\partial \theta^{i}$ denotes a partial derivative with respect to $\theta^{i}$. Now consider a small perturbation around the true parameters $\boldsymbol{\theta}_{\text{bf}} = \boldsymbol{\theta}_{0} + \Delta \boldsymbol{\theta}$ for $\Delta\boldsymbol{\theta} = (\boldsymbol{\theta}_{\text{bf}} - \boldsymbol{\theta}_{0})$. By applying the linear signal approximation, an expansion in $\Delta\boldsymbol{\theta} \ll 1$ to first order in our model templates gives
\begin{equation}
h^{(X)}_{m}(\boldsymbol{\theta}_{0}) \approx h^{(X)}_{m}(\boldsymbol{\theta}_{\text{bf}}) - \partial_{i}h^{(X)}_{m}(\boldsymbol{\theta}_{\text{bf}})\Delta\theta^{i}\, , \label{eq:LSA}
\end{equation}
From this point onward, we will drop the (fixed) time coordinate $t$ for notational convenience. Equation \eqref{eq:LSA} can then be used in the expression $d^{(X)} - h^{(X)}_{m}$ to find 
\begin{widetext}
\begin{align}
    d^{(X)} - h^{(X)}_{m}(\boldsymbol{\theta}_{\text{bf}}) & = n^{(X)} + h^{(X)}_{e}(\boldsymbol{\theta}_{0}) - h^{(X)}_{m}(\boldsymbol{\theta}_{\text{bf}}) \nonumber \\ 
    & = n^{(X)} + h^{(X)}_{e}(\boldsymbol{\theta}_{0}) - h^{(X)}_{m}(\boldsymbol{\theta}_{0}) + h^{(X)}_{m}(\boldsymbol{\theta}_{\text{0}})  - h^{(X)}_{m}(\boldsymbol{\theta}_{\text{bf}}) \nonumber \\
    & \approx n^{(X)} + \delta h^{(X)}(\boldsymbol{\theta}_{0}) - \partial_{i}h^{(X)}_{m}(\boldsymbol{\theta}_{\text{bf}})\Delta\theta^{i} + (\Delta\theta^{i})^2 \label{eq:LSA_application_dmh}
\end{align}
for $\delta h^{(X)}(\boldsymbol{\theta}_{0}) = h_{e}^{(X)}(\boldsymbol{\theta}_{0}) - h_{m}^{(X)}(\boldsymbol{\theta}_{0})$ denoting residuals between the true waveform $h_{e}^{(X)}$ and the approximate waveform $h_{m}^{(X)}$. When there are no mismodelling errors present, the term $\delta h^{(X)} = 0$. Substituting equation \eqref{eq:LSA_application_dmh} into \eqref{eq:max_likelihood}, one obtains at first order in $\Delta \theta^{i}$
\begin{equation}\label{eq:big_nasty_equation}
\sum_{X = \{A,E,T\}}[\Delta\theta^{j}(\partial_{j}h_{m}(\boldsymbol{\theta}_{\text{bf}})|\partial_{i}h_{m}(\boldsymbol{\theta}_{\text{bf}}))^{(X)} - (\partial_{j}h_{m}(\boldsymbol{\theta}_{\text{bf}}) | n)^{(X)} - (\partial_{j}h_{m}(\boldsymbol{\theta}_{\text{bf}}) | \delta h(\boldsymbol{\theta}_{\text{0}}))^{(X)}] = 0 
\end{equation}
Defining the matrix
\begin{equation}\label{eq:FM_AET}
(\Gamma_{AET})_{ij} =
\sum_{X = \{A,E,T\}}(\partial_{i}h_{m}(\boldsymbol{\theta}_{\text{bf}})|\partial_{j}h_{m}(\boldsymbol{\theta}_{\text{bf}}))^{(X)}
\end{equation}
it is then possible to invert the matrix-vector equation \eqref{eq:big_nasty_equation} to calculate $\Delta \theta^{i}$
\begin{equation}
\Delta \theta^{i} = (\Gamma_{AET}^{-1})^{ij}\left[\sum_{X}(\partial_{j}h_{m}(\boldsymbol{\theta}_{\text{bf}})|n)^{(X)} +(\partial_{j}h_{m}(\boldsymbol{\theta}_{\text{bf}})|\delta h(\boldsymbol{\theta}_{\text{0}}))^{(X)} \right]\label{eq:overall_FM_bias}.
\end{equation}
\end{widetext}

In the presence of noise fluctuations $n^{(X)}$ and waveform modelling errors $h_{e} \neq h_{m}$, there are two sources of discrepancy between the recovered parameters $\theta^{i}_{\text{bf}}$ and true parameters $\theta^{i}_{0}$ described by equation \eqref{eq:overall_FM_bias}. Each of these terms represent a \emph{statistical error}, determined by the presence of noise and the second a \emph{systematic error}, a consequence of non-faithful model templates $h_{m}$
\begin{align}
    \Delta \theta^{i}_{n} &= (\Gamma^{-1}_{AET})^{ij}\sum_{X}(\partial_{j}h_{\text{m}}(\boldsymbol{\theta}_{\text{bf}})|n)^{(X)}, \label{eq:noise_fluc}\\
    \Delta \theta^{i}_{\text{sys}} &= (\Gamma^{-1}_{AET})^{ij}\sum_{X}(\partial_{j}h_{\text{m}}(\boldsymbol{\theta}_{\text{bf}})|\delta h(\boldsymbol{\theta}_{0}))^{(X)}. \label{eq:sys_bias}
\end{align}

The first term \eqref{eq:noise_fluc} enforces a statistical fluctuation to the recovered parameters. Since the noise has mean zero, the statistic $\theta^{i}_{\text{n,bf}} = \theta^{i}_{0} + (\widehat{\Delta \theta^{i}_{n}})_{AET}$ is an unbiased estimator of the true parameters such that $\mathbb{E}[\theta^{i}_{\text{n,bf}}] = \theta^{i}_{0}$. The quantity \eqref{eq:sys_bias} governs the bias in the recovered parameters due to using inaccurate waveform models where $\delta h^{(X)}(\boldsymbol{\theta}_{0}) \neq 0$. The overall bias is thus given by the second term in equation \eqref{eq:sys_bias}
\begin{equation}\label{eq:bias_mismodelling_expectation}
\mathbb{E}[\theta^{i}_{\text{bf}}] = \theta_{0}^{i} +  (\Gamma^{-1}_{AET})^{ij}\sum_{X}(\partial_{j}h_{\text{m}}(\boldsymbol{\theta}_{\text{bf}})|\delta h(\boldsymbol{\theta}_{0})^{(X)}). 
\end{equation}

Note that since $h \sim \rho$ and $S_{n}(f) \sim \rho^{0}$, we have that $\Gamma^{-1} \sim \rho^{-2}$ giving the scaling relationship for the statistical uncertainty $(\widehat{\Delta \theta^{i}_{n}})_{AET} \sim O(1/\rho)$. Similarly, for the systematic error, the scaling relationship is given by $(\widehat{\Delta \theta^{i}_{\text{sys}}})_{AET} \sim O(\rho^{0})$ and is thus independent of the SNR of the source. Therefore, if the signal-to-noise ratio of the underlying signal is large, then the (relative) magnitude of the systematic error will be larger when compared to the statistical fluctuation. Further details can be found in \cite{Flanagan:1997kp, Miller2005, Cutler:2007mi, antonelli2021noisy}.

In the limit of small waveform modelling errors $|\delta h^{(X)}|^{2} \ll \rho^{0}$ and high SNR, the Fisher matrix yields an approximation to the predicted covariance matrix of the posterior distribution
\begin{equation}\label{eq:inverse_FM}
    \mathbb{E}[(\Delta\theta^{i}_{\text{bf}})(\Delta\theta_{\text{bf}}^{j})] \approx  (\Gamma_{AET}^{-1})^{ij}\,.
\end{equation}
with rooted diagonal elements an approximation to how well one can constrain the parameters of the system. The Fisher matrix is widely used within gravitational-wave astronomy to cheaply compute precision measurements on parameters of interest. Precision measurements are given by the rooted diagonals of the inverse of the Fisher matrix 
\begin{equation}\label{eq:stat_error}
\Delta \theta^{i}_{\text{stat}} = \sqrt{(\Gamma^{-1})^{ii}} \quad \text{no sum.}
\end{equation}
where $\Delta \theta^{i}_{\text{stat}}$ is the statistical error, the $1\sigma$ deviation (through expectation) in the recovered parameters due to noise fluctuations. For systematic studies in Cutler-Valisneri framework~\cite{Cutler:2007mi}, the ratio between \eqref{eq:sys_bias} and \eqref{eq:stat_error} is computed. If the bias on parameters exceed the statistical uncertainty, then the proposed waveform model is not suitable for parameter estimation. 

In our work, we will not focus on error induced due to approximate waveform models. Instead, we will focus on the notion of \emph{coverage} given by an approximate posterior distribution. The ``coverage'' of a posterior density describes the probability that the true parameters are contained within the posteriors credible set. The Cutler Valisneri formalism can be discussed in terms of coverage of a posterior: If the (assumed normal) approximate posterior has a 68\% coverage, then the waveform model will be deemed suitable for parameter estimation\footnote{In other words, for an approximate one dimensional Gaussian distribution, if the true parameters are contained within the 68\% level credible interval $[\hat{\mu} - \hat{\sigma}, \hat{\mu} + \hat{\sigma}]$, the model is deemed suitable for parameter estimation. Here hatted quantities are estimates of the posterior means and standard deviations.}. The primary focus of our work is to \emph{calibrate} the coverage of an approximate posterior to a much higher level, specified by the user. The final (calibrated) credible region will then contain the true parameters with higher level of probability the original credible region. In other words, the coverage of the new calibrated approximate posterior will be larger, indicating greater certainty that we have recovered the correct parameters to a reasonable certainty. To discuss this in more detail, it is necessary to introduce the Bayesian tools we will use to perform parameter estimation.

The Fisher matrix formalism leading to equations(\ref{eq:noise_fluc}, \ref{eq:sys_bias}, \ref{eq:inverse_FM}) can be used to compute precision measurements on parameters and potential sources of bias on the recovered parameters $\boldsymbol{\theta}_{\text{bf}}$. However, given an observed data stream $d_{o}$, we wish to make inference on parameters $\boldsymbol{\theta}$ that govern the structure of the underlying data set (and thus the GW signal). 

\subsection{Bayesian theory}

The standard procedure used within GW astronomy to estimate parameters of a signal $h^{(X)}$ given observation of a set of data streams $d_{o}^{(X)}$ is Bayesian inference. At the heart of Bayesian theory lies Bayes' theorem: 
\begin{eqnarray}
    p(\boldsymbol{\theta}|d_{o}) &=& \frac{p(d_{o}|\boldsymbol{\theta})p(\boldsymbol{\theta})}{p(d_o)} \\
    &\propto& p(d_{o}|\boldsymbol{\theta})p(\boldsymbol{\theta}),
\end{eqnarray}
where $p(\boldsymbol{\theta}|d_{o})$ is the posterior density of unknown parameters $\boldsymbol{\theta}$ given the observation of a data stream $d_{o}$, $p(d_{o}|\boldsymbol{\theta})$ the likelihood function and $p(\boldsymbol{\theta})$ is the prior distribution, reflecting our beliefs on parameters $\boldsymbol{\theta}$ before observing the data. The marginal likelihood $p(d_o) = \int_{\boldsymbol{\theta}\in\boldsymbol{\Theta}}p(d_{0}|\boldsymbol{\theta})p(\boldsymbol{\theta})\,\text{d}\boldsymbol{\theta}$ is a constant over the parameter space and is unnecessary for our work. 

Stochastic sampling algorithms, such as Markov chain Monte Carlo (MCMC), are used to obtain random samples $\boldsymbol{\theta}$ from the posterior density $p(\boldsymbol{\theta}|d_o)$ by constructing a Markov chain whose steady-state distribution is the posterior distribution of interest.  The posterior distribution is then summarised using Monte Carlo integration to compute moments such as the posterior mean $\mathbb{E}_{p(\boldsymbol{\theta}|d_o)}[\boldsymbol{\theta}]$ or quantifying levels of precision on how well we can constrain parameters. In this work, we use the MCMC ensemble sampler \texttt{emcee}~\cite{emcee_hammer} to obtain samples from $p(\boldsymbol{\theta}|d_o)$.

To obtain the \emph{exact} posterior density, one would use the likelihood function \eqref{eq:whittle_likelihood_signal} with model templates $h_{\text{m}}$ precisely equal to the true waveform within the data stream $h_{\text{e}}$. This would yield an unbiased result in the recovered parameters, a consequence of generating an \emph{exact} posterior density $p(\boldsymbol{\theta}|d_o)$. However, in the context of gravitational-wave astronomy, this is unfeasible. The two-body problem in general relativity has no exact solution, and the most numerically accurate are NR waveforms~\cite{boyle2019sxs} that are computationally prohibitive for MCMC algorithms. Instead, we must make do with approximate models that are both \emph{fast} to generate and \emph{faithful} with their true counterpart. Hence, all cases of gravitational-wave astronomy, we in fact sample from an approximate posterior density $\tilde{p}(\boldsymbol{\theta}|d_o)$:
\begin{equation}
    \tilde{p}(\boldsymbol{\theta}|d_o) \propto \tilde{p}(d_o|\boldsymbol{\theta})p(\boldsymbol{\theta}),
\end{equation}
with $h_{m} \neq h_{e}$ in the approximate likelihood function \eqref{eq:whittle_likelihood_signal}. As discussed in section \ref{sec:FM}, this would result in a biased set of parameters. 

Generating samples from an approximate posterior distribution is very common in Bayesian inference \citep{pritchard:1999, besag:1975, wood:2010, jordan:1999}. When an approximate likelihood $\tilde{p}(d_o|\boldsymbol{\theta})$ is used, the resulting posterior inference will be distorted\footnote{By distortion we refer to the non-zero statistical distance between two distributions, say $p_{1}$ and $p_{2}$. For example, such a distortion (and thus statistical distance) could be measured by the Kullback-Leibler (KL) divergence.} with respect to the exact posterior distribution $p(\boldsymbol{\theta}|d_o)$. Within the statistical literature, various estimators have been proposed to measure this distortion \citep{fearnhead:2012, prangle:2014, yao:2018}. Most of the existing methods are based on \citep{geweke:2004} and \citep{cook:2006}. Their original motivation was to check 
for correct sampling from the posterior distribution  based on the following equality for all $\boldsymbol{\theta},\boldsymbol{\theta}'\in\Theta$:
\begin{equation}
    \mathbb{E}_{d_o,\boldsymbol{\theta}'} [p(\boldsymbol{\theta}|d_o)] =p(\boldsymbol{\theta}),
\end{equation}
i.e.\ the integral of the exact posterior density with respect to the generative model $p(d_o|\boldsymbol{\theta}')p(\boldsymbol{\theta}')$ is equal to the prior. 
Then they constructed statistical tests to check the validity of this relationship when replacing $p(\boldsymbol{\theta}|d_o)$ by an approximate posterior density
$\tilde{p}(\boldsymbol{\theta}|d_o)$. However, these types of tests can falsely accept the hypothesis of correct sampling even if the approximate likelihood is far from the exact likelihood, see e.g.\
\citep{lee:2019,prangle:2014}. Moreover, they do not quantify the distortion for a particular observation. 

\citep{lee:2019} showed how to quantify the distortion of an approximate posterior credible interval conditional on the observed data by estimating its operational coverage as described in Sections~\ref{subsec:idealoperationalcoverage} and \ref{subsec:practicaloperationalcoverage}. We present a practical operational coverage estimator for gravitational-wave problems in Section~\ref{subsec:practicaloperationalcoverage} and how to calibrate approximate credible set in Section~\ref{subsec:Credible interval calibration via operational coverage estimation}.

% III %%%%%%%%%%%%%%%%%%%%%%%%%%%%%%%%%%%%%%%%%%%%%%%%%%%%%%%%%%%%%%%%%%%%%%%%%%%%%

\section{General calibration methodology}\label{sec:general_calibration_methodology}

\subsection{Ideal operational coverage estimation}\label{subsec:idealoperationalcoverage}

Let $d_o$ represent the observed data. When we refer to ``coverage," we are describing the posterior probability that a credible set, determined by a specific prior and likelihood function forming the posterior, contains the true parameters we intend to estimate.

Let $\widetilde{C}_{d_o}$ and $C_{d_o}$ be the level $\alpha$ posterior credible sets calculated using $\tilde{p}(\boldsymbol{\theta}|d_o) \propto \tilde{p}(d_o|\boldsymbol{\theta})p(\boldsymbol{\theta})$
    and $p(\boldsymbol{\theta}|d_o) \propto p(d_o|\boldsymbol{\theta})p(\boldsymbol{\theta}),$ respectively, i.e.,
\begin{eqnarray}\label{credibleset}
\alpha&=& P(\boldsymbol{\theta}\in C_{d_o})=\displaystyle\int \mathbbm{1}_{ C_{d_o}}(\boldsymbol{\theta})p(\boldsymbol{\theta}|d_o) \text{d}\boldsymbol{\theta} \\
\alpha&=& \widetilde{P}(\boldsymbol{\theta}\in \widetilde{C}_{d_o})=\displaystyle\int \mathbbm{1}_{ \widetilde{C}_{d_o}}(\boldsymbol{\theta})\tilde{p}(\boldsymbol{\theta}|d_o) \text{d}\boldsymbol{\theta} \nonumber
\end{eqnarray} 

where $\mathbbm{1}$ denotes the indicator function, i.e.\ $\mathbbm{1}_A(x)=1$ if $x\in A$ and 0 otherwise. 
A coverage of $\alpha$ is only guaranteed if the data is distributed according to the assumed generative model. This means that if the data $d_o$ is actually generated from the specified likelihood $p(d_o|\boldsymbol{\theta})$, then  $C_{d_o}$ achieves the nominal level $\alpha$.

However, since the likelihood $\tilde{p}(d_o|\boldsymbol{\theta})$ of the approximate posterior does not correspond to the generative model, its level $\alpha$ credible set $\widetilde{C}_{d_o}$ does not achieve the nominal level $\alpha$ but only an \textit{operational} coverage probability
 
\begin{equation}\label{eq:operational} b(d_o)= P(\boldsymbol{\theta} \in \widetilde{C}_{d_o}) = \displaystyle\int \mathbbm{1}_{ \widetilde{C}_{d_o}}(\boldsymbol{\theta})p(\boldsymbol{\theta}|d_o) \text{d}\boldsymbol{\theta} \end{equation}
that is not generally equal to the nominal coverage, $\alpha$.  

If $b(d_o)\gg \alpha$ or $b(d_o)\ll \alpha$, this would indicate a poor approximation. Thus $|\alpha-b(d_o)|$ measures the distortion or discrepancy in coverage of the credible set at the observed data $d_o$. If an approximation is not good enough for a user, there are two possible approaches to fix this: using a different approximation, such as using a more accurate but more costly waveform model, or correcting the posterior itself \citep{xing:2020}.

In practice, we often generate samples from the approximate posterior $\tilde{p}(\boldsymbol{\theta}|d_o)$ using MCMC and estimate $\widetilde{C}_{d_o}$. If we denote this estimator of $\widetilde{C}_{d_o}$ by $\widehat{C}_{d_o}$, we get the \textit{realised operational coverage} probability
\begin{equation}\label{eq:realised} 
b_r(d_o)= P(\boldsymbol{\theta} \in \widehat{C}_{d_o}). \end{equation}
 
The realised operational coverage probability in Eq.~(\ref{eq:realised}) can be estimated using the standard Monte Carlo method, i.e., sampling  from the exact posterior $p(\boldsymbol{\theta}|d_o)$ and taking the proportion of the samples that are inside credible intervals $\widehat{C}_{d_o}$. This estimate takes the Monte Carlo error of estimating the credible set into account. However, this procedure will not be practical because it needs samples from the exact posterior $p(\boldsymbol{\theta}|d_o)$, which may be expensive and impractical to sample from.  An example here would be generating an exact posterior density $p(\boldsymbol{\theta}|d)$ using the most accurate, but computationally prohibitive numerical relativity waveforms for MBHs. Instead, using techniques from regression, we show that it is possible to provide operational coverage estimators without sampling from the exact posterior distribution $p(\boldsymbol{\theta}|d)$ in the next section. 

\subsection{Practical operational coverage estimation}\label{subsec:practicaloperationalcoverage}

Operational coverage estimators that do not require simulation from the exact posterior have been suggested \citep{lee:2019, xing:2019}. These are based on logistic regression (binary classification) and (annealed) importance sampling. In this paper, we use the logistic regression as the operational coverage.

The set up is the following. For $j=1,\ldots,J$, we sample $\boldsymbol{\theta}_{(j)}$ from the prior, $\boldsymbol{\theta}_{(j)} \sim p(\boldsymbol{\theta})$ and generate data $d_{(j)}= h_{\text{e}}(\boldsymbol{\theta}_{(j)}) + n$. For each $d_{(j)}$, we estimate a credible set $\widehat{C}_{d_{(j)}}$ of $\tilde{p}(\boldsymbol{\theta}_{(j)}|d_{(j)})$ and, $c_j=\mathbbm{1}(\boldsymbol{\theta}_{(j)} \in \widehat{C}_{d_{(j)}})$ which is regarded as a  Bernoulli trial with success probability $b_r(d_{(j)})$ associated with the data itself, i.e. 
\[ c_j \sim \mbox{Bernoulli}(b_r(d_{(j)}))\,.\] 
If we can fit a logistic regression to $c_j$ with $d_{(j)}$ as predictors, one can use the model to predict $b_r(d_o)$ with the observed data $d_o$ . We denote this prediction as $\bar{b}_r(d_o)$. In literature, a semiparametric regression (GAM) \citep{lee:2019} and a Bayesian Additive Regression Tree (BART) \citep{xing:2019} were used.

 Theoretical properties of traditional parametric and nonparametric regression often assume that the number of samples is larger than the dimension of predictor i.e., $|d| \ll J$. As often the case within gravitational-wave data analysis, this assumption is violated when the length of $d_{(j)}$ is larger than the training data set size $J$.
 In this paper, we use an artificial neural network (ANN) with the sigmoid activation function to fit the binary classification for practicality. To enhance the efficiency of the artificial neural network (ANN) \citep{Hintonand:2006,Fournier:2019}, we utilize an autoencoder to project $d_{(j)}$ into a lower-dimensional subspace while capturing the main features of the data \citep{Quentin2019}. The autoencoder comprises two neural networks: an encoder network that maps the input data into a lower-dimensional latent space, and a decoder network that recreates the input from the encoded representation.

For realistic applications of data analysis, the choice of the nominal value $\alpha$ is chosen by the user. The proposed estimators by \citep{lee:2019,xing:2019} are conditioned on a nominal value and restrict the operational coverage estimate to a particular nominal coverage. In an attempt to tackle this issue, the nominal level $\alpha$ is also taken as an input to fit the classification. The training set is $\{d_{(j)},c_j,\alpha_j\}^J_{j=1}$ where $\alpha_j\sim p(\alpha)$, $d_{(j)}= h_{\text{e}}(\boldsymbol{\theta}_{(j)}) + n$ and $c_j= \mathbbm{1}_{\widehat{C}_{d_{(j)},\alpha_j}}(\boldsymbol{\theta}_{(j)})$. Here, $\widehat{C}_{d_{(j)},\alpha_j}$ is an estimated credible set of $\tilde{p}(\boldsymbol{\theta} |d_{(j)})$ with a nominal level, $\alpha_j$. i.e., $\alpha_j = \int \mathbbm{1}_{\widehat{C}_{d_{(j)},\alpha_j}} (\boldsymbol{\theta}) \tilde{p}(\boldsymbol{\theta} |d_{(j)}) d\boldsymbol{\theta} $. 

\begin{figure*} 
\includegraphics[width = 16cm, height = 8cm]{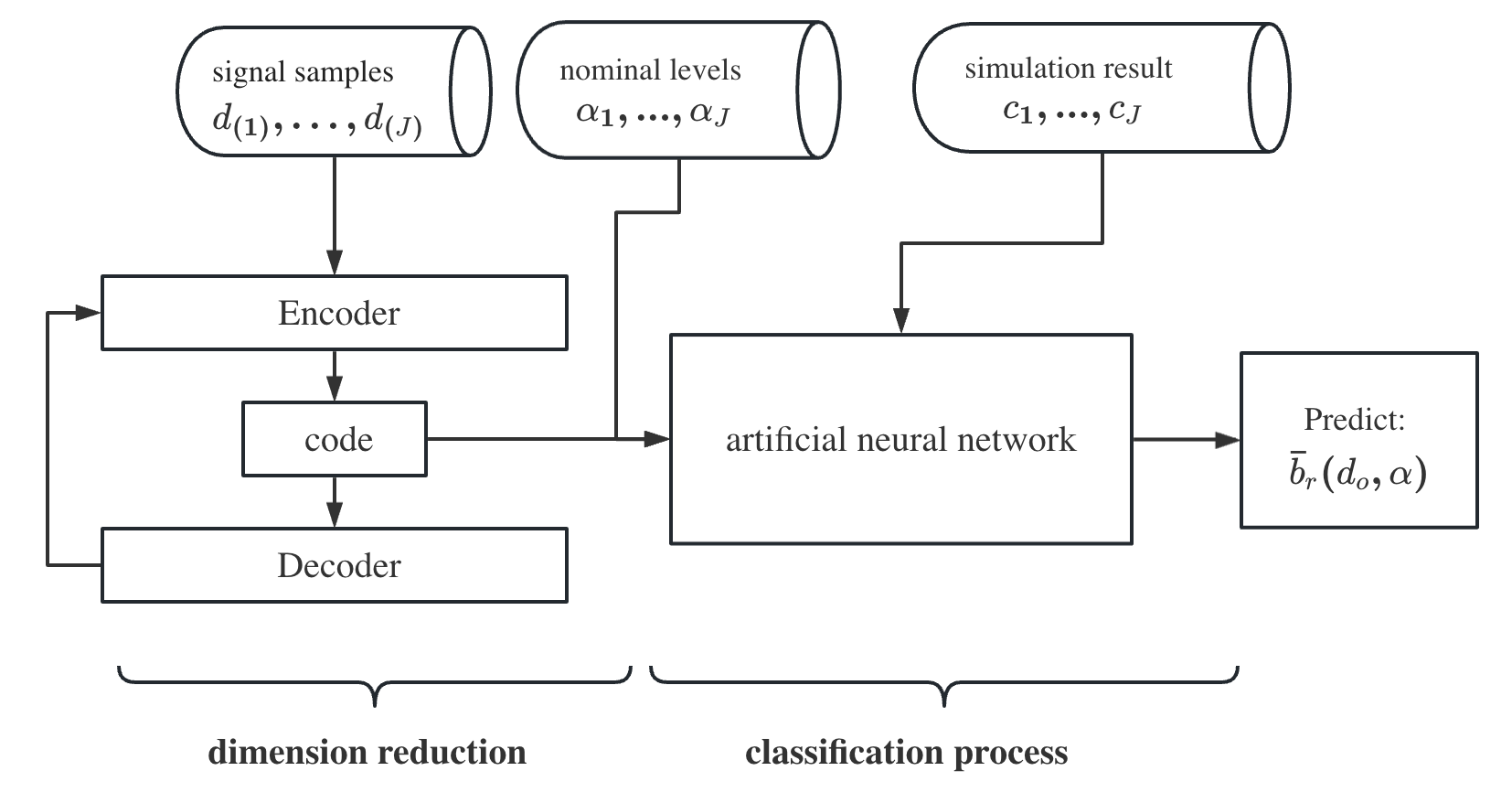} 
\caption{The flow chart describing the overall procedure of our operational coverage estimator. 
For $j=1,...,J$, generate $\boldsymbol{\theta}_{(j)} \sim p(\boldsymbol{\theta})$ and data streams generated using an exact model with noise $d_{(j)} = h^{(X)}_{e}(t;\boldsymbol{\theta}_{(j)}) + n^{(X)}(t)$ are passed into an encoder where dimensional reduction is applied. After this process, nominal levels are simulated $\alpha_j \sim p(\alpha)$ and $\widehat{C}_{d_{(j)},\alpha_j}$ of an approximate posterior distribution using the Fisher based parameter estimation scheme returns $c_j$, where $c_j = 1$ if $\boldsymbol{\theta}_{(j)} \in \widehat{C}_{d_{(j)},\alpha_j}$ and 0 otherwise. 
The classifier is subsequently trained, enabling us to predict $\bar{b}_r(d_o,\alpha)$ at a specified nominal level $\alpha$.} 
\label{fig:process}  \end{figure*}

Figure \ref{fig:process} shows the procedure for constructing the operational coverage estimator. It consists of two components, dimensional reduction and classification. First, we train the encoder function in the autoencoder to reduce the dimension of $\{d_{(j)}\}_{j=1}^J$. Then the compressed data and nominal values are fed to an ANN as an input feature. To obtain the target output $c_1,...,c_J$,  the classifier is trained. The operational coverage, which is the success probability of the fitted classifier, can be predicted with $d_o$ at a desired nominal level of $\alpha$, and it is denoted by $\bar{b}_{r}(d_o,\alpha)$.

We should point out that data stream $d_{(j)}$, as an input of the proposed classifier, can be either an actual signal ($h_e+n$) or discrete Fourier transform of signal ($\hat{h}_e+\hat{n}$). Alternatively, an adequate summary of signals can be used, and it is denoted by $S(d_{(j)})$. If a summary is used, the training set is $\{S(d_{(j)}),c_j,\alpha_j\}^J_{j=1}$ and, the prediction is made with a nominal level of $\alpha$ and the summary of the observed data, $\bar{b}_r(S(d_o))$. Our choice of the summary for numerical simulation studies are included in Section \ref{sec:toy_model} and \ref{sec:massive_black_holes_application}.

We use $k$-fold cross-validation to predict success probabilities $\bar{b}_r(d_o)$ and class at $d_o$, but the uncertainty of the operational coverage estimate is not attainable (i.e., only a point estimate is available with this approach). For a simple logistic regression, the Delta method \citep{xu2005using} and the bootstrap method \citep{efron1986bootstrap} are therefore applied to measure the prediction errors in Section \ref{sec:toy_model}. We emphasize here that there does not exist any unbiased and universal estimator of the variance of $k$-fold cross-validation that is valid under all distributions \citep{Bengio2004}.

\subsection{Credible interval calibration via operational coverage estimation}\label{subsec:Credible interval calibration via operational coverage estimation}

There are no formal guidelines on how to interpret operational coverage. One may consider correcting an approximate posterior distribution. Within the Approximate Bayesian Computation (ABC) framework, adjustment procedures for bias and frequentist coverage of Bayesian credible sets were suggested when the credible set does not have the correct nominal coverage probability in the frequentist sense \citep{menendez:2014}. With a focus on coverage in the Bayesian sense, a distortion map estimator using machine learning methods was proposed by \citep{xing:2020}, and is limited to one-dimensional problems. 

The application of an operational coverage estimator enables us to determine the approximate credible set level that yields the desired posterior coverage level. Since an exact inverse map of the trained ANN is not always feasible or necessary, the simplest way is to read off from a plot of operational coverage estimates against nominal values at observed signal $d_{o}$.

\begin{figure*} 
\includegraphics[width=16cm,height=6cm]{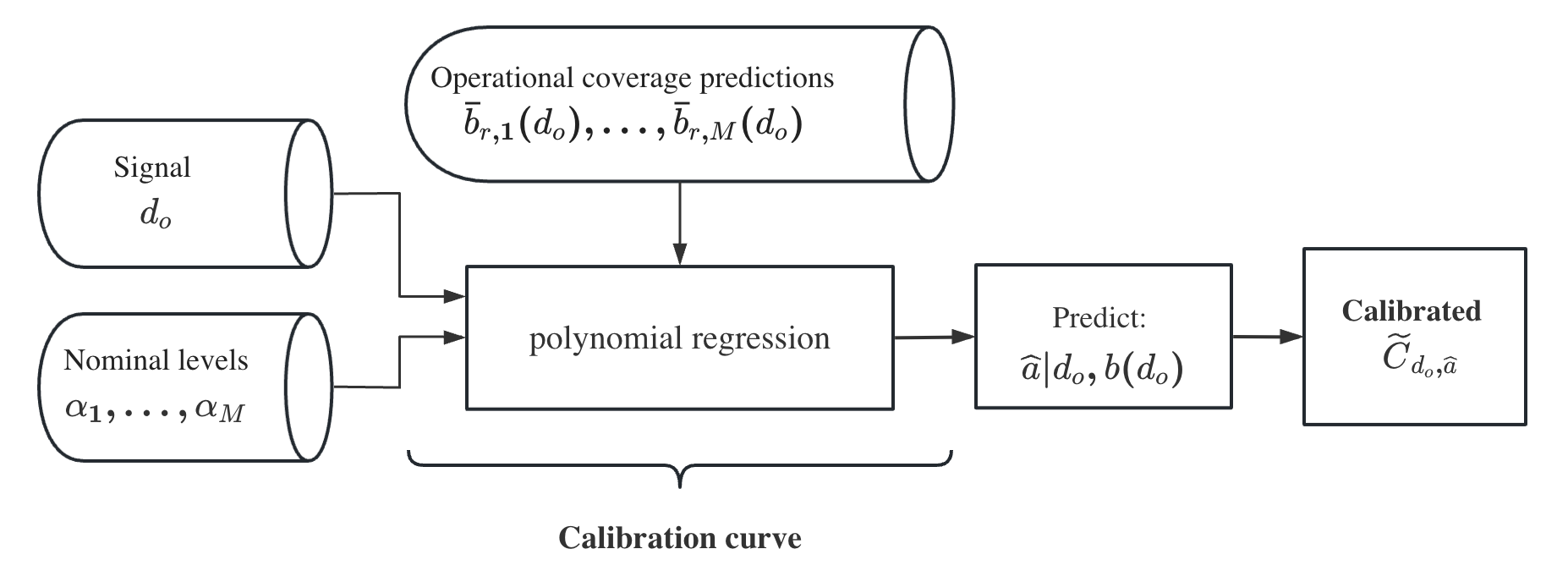} 
\caption{The flow chart describes the overall procedure of the calibration curve. After the procedure given in Figure \ref{fig:process} is completed, for a nominal level sample $\alpha_j \sim p(\alpha)$ the operational coverage is predicted $\bar{b}_{r,j}(d_o)$ using the estimator, $j=1,...,M$. Taking $\bar{b}_{r,j}(d_o)$ as a predictor and $\alpha_j$ as a response, polynomial regression is fitted then the calibrated nominal level $\widehat{\alpha}$ is predicted at the desired operational coverage $b(d_o)$. } 
\label{fig:inverse}  \end{figure*}

We construct the calibration curve for an observed signal $d_{o}$ that the nominal level is parameterized by operational coverage and the procedure is summarized in Figure \ref{fig:inverse}. The training data $\{\alpha_j,\bar{b}_{r,j}(d_o)\}_{j=1}^M$ is constructed by finding the operational coverage at each of the $M$ nominal levels in $ [\alpha_{\min},\alpha_{\max}]$ 
from the estimator (Figure \ref{fig:process}). Taking a nominal level ($\alpha_j$) as a response and operational coverage ($\bar{b}_{r,j}(d_o)$) as a predictor, the $K$-degree polynomial regression is fitted.  The value for $K$ is chosen by minimizing the residual mean squared error, i.e., 
$$\alpha_j = c_0 + \sum_{k=1}^K c_k \bar{b}_{r,j}(d_o)^k\,, \hspace{.5cm} j=1,...,M\,.$$

Given the desired operational coverage $b(d_o)$, the \textit{calibrated nominal coverage} level, $\widehat{\alpha}$, is estimated from the calibration curve for $d_o$ and, with the calibrated approximate credible set $\widehat{C}_{d_o,\hat{\alpha}}$,  $P(\theta \in \widehat{C}_{d_o,\hat{\alpha}}) = b(d_o)$. 

\bigskip 

We demonstrate how to estimate an operational coverage and calibrate a credible set using a simple toy example in Section \ref{sec:toy_model} and a massive black hole problem in Section \ref{sec:massive_black_holes_application}.

%%%%%%%%%%%%%%%%%%%%%%%%%%%%%%%%%%%%%%%%%%%%%%%%%%%%%%%%%%%%%%%%%%%%%%%%%%%%%%%%%%%%%%%

\section{Application: Simple toy model}\label{sec:toy_model}

In this section, we present a simple toy model to illustrate our calibration procedure discussed in section \ref{sec:general_calibration_methodology}. 

\subsection{Set up and Fisher matrix validation}

In this section, we will consider a data stream of the form
\begin{equation}\label{eq:data_stream_toy}
d(t) = h_{e}(t;\boldsymbol{\theta}) + n(t)
\end{equation}
with an exact template of the form 
\begin{equation}\label{eq:exact_toy_model}
h_{e}(t;a_{0},f_{0},\dot{f}_{0}) = a_{0}\sin \left(2\pi t\left[f_{0} + \frac{1}{2}\dot{f}_{0}t\right]\right).
\end{equation}
Here, we have the true values for parameters $\boldsymbol{\theta}_0 = \{a_{0} = 5\cdot 10^{-21}, f_{0} = 10^{-3}\,\text{Hz}, \dot{f}_{0} = 10^{-8}\,\text{Hz}/\text{s}\}$.

We use model templates
\begin{equation}\label{eq:simple_waveform_model}
h_{m}(t;a,f,\dot{f},\epsilon) = a \sin \left(2\pi t\left[f + \frac{1}{2}\dot{f}t\right](1 - \epsilon)\right).
\end{equation}

Here $\epsilon \ll 1$ is used as a tuneable parameter allowing deviations from the exact model \eqref{eq:exact_toy_model} $h_{\text{e}}(t;\boldsymbol{\theta},\epsilon = 0)$ given by an approximate model $h_{\text{m}}(t;\boldsymbol{\theta}, \epsilon \neq 0)$. For simplicity, we do not take into account the response function through TDI as outlined in section \ref{sec:GW_data_analysis}. Hence, when calculating the inner product \eqref{eq:inner_prod}, which is used in likelihood~\eqref{eq:whittle_likelihood_noise}, Fisher matrix~\eqref{eq:FM_AET} and SNR~\eqref{eq:SNR_X} calculations, we only consider a single data stream and use the approximate LISA-like PSD defined by Equation (1) of~\cite{robson2019construction}.

Using equations \eqref{eq:real_part_noise} and \eqref{eq:imag_part_noise}, we generate stationary Gaussian noise in the frequency domain to construct the toy model data stream \eqref{eq:data_stream_toy}. In this example, we will set $\epsilon = 10^{-6}$, allowing for a discrepancy between the exact model $h_{e}$ and approximate model $h_{m}$. For $\epsilon = 0$ and over a time of observation of 30 hours sampled with cadence $\Delta t \sim 200\,$ seconds, we observe an optimal matched filtering SNR $\rho \sim 188$. The length of the data stream is $N = 2^{16}$.

Our calibration technique requires multiple parameter estimation simulations using an approximate model to then estimate the operational coverage. As discussed in section~\ref{sec:intro}, this is extremely computationally intensive so we approximate samples from the approximate posterior density using a Fisher matrix approach instead.

To validate our Fisher matrix approach, we inject an exact model with $\epsilon = 0$ and recover with an approximate model with $\epsilon = 10^{-6}$ using the \texttt{emcee} algorithm. We generate 31,000 samples from the approximate posterior under $h_{m}(t;\boldsymbol{\theta}),\epsilon = 10^{-6})$, and then discard 6,000 samples as burn-in. In parallel, we compute the Fisher matrix \eqref{eq:FM_AET} and then sample from a multivariate Gaussian 
\begin{align}\label{eq:sample_multivariate_gaussian_FM}
\begin{split}
\boldsymbol{\theta} &\sim \mathcal{N}\left(\boldsymbol{\theta}_{0} + \boldsymbol{\theta}_{\text{bias}}, \Gamma^{-1}(h_{m})\right)\, ,\\
\theta^{i}_{\text{bias}} &=  [\Gamma^{-1}(h_{m})]^{ij}(\partial_{j}h_{\text{m}}|\delta h + n)\,,
\end{split}
\end{align}
with $\theta^{i} \in \boldsymbol{\theta}_{\text{bias}}$. Here equation~\eqref{eq:sample_multivariate_gaussian_FM} is evaluated at the true parameters $\boldsymbol{\theta}_{0}$ and $\delta h = h_{e}(t;\boldsymbol{\theta}_{0}) - h_{m}(t;\boldsymbol{\theta}_{0},\epsilon = 10^{-6})$. We plot the approximate posterior densities in Figure \ref{transformationg:3dFMcheck}. The blue curve is generated via MCMC and the green curve generated via the Fisher matrix computation. This simulation has shown that the Fisher matrix can be used as a suitable approximation to the posterior density. Having verified the Fisher matrix is a suitable approximation, we then use it to generate approximate posteriors in bulk in order to apply the calibration procedure discussed in \ref{sec:general_calibration_methodology}. This is the focus of the next section.

\subsection{Calibration procedure}\label{sec:calibration_procedure_toy}

The calibration technique described in section \ref{sec:general_calibration_methodology} requires a training set, built from the prior space of samples and the resultant generation of a family of approximate posteriors. We first focus our attention on a single-parameter study, then generalise to the three-parameter study at the end of this section. 

For a single-parameter study, $\dot{f}$ is unknown and the true values for $a,f$ are used. i.e., $\boldsymbol{\theta}=\dot{f}$. The uniform prior is assigned for $\dot{f}$,
\begin{equation}\label{pr_fdot}
\dot{f} \sim \text{U}[\dot{f}_{0}-10^{-13}, \dot{f}_{0}+10^{-13}]\,\text{Hz/s} \,.
\end{equation}

The training data $\{S(d_{(j)}), c_j,\alpha_j\}^{5000}_{j=1}$ is generated to obtain a practical operational coverage estimator. For $j=1,...,5000$, $\boldsymbol{\theta}_{(j)}$ is generated from the prior (\ref{pr_fdot}) and $\alpha_j\sim U[0.78,0.97]$. For each of prior sample $\boldsymbol{\theta}_{(j)}$, the data stream $d_{(j)} = \hat{h}_{e,(j)}(f;\boldsymbol{\theta}_{(j)}) + \hat{n}_{(j)}$ is generated by adding a noise $\hat{n}_j$ through \eqref{eq:discrete_noise_gen} to the exact reference signal $\hat{h}_{e,(j)}$. An approximate posterior density is in the form of a multivariate Gaussian \eqref{eq:sample_multivariate_gaussian_FM} using the inverse Fisher matrix \eqref{eq:inverse_FM} and expectation for the bias \eqref{eq:overall_FM_bias} at $\boldsymbol{\theta}_{(j)}$. An output $c_j$ is obtained from the $\alpha_j$ credible set of the approximate posterior density. Instead of $d_{(j)}$, the real part of discrete Fourier transform of signal is used to find the practical operational coverage estimator. i.e, $S(d_{(j)}) = Re(d_{(j)})$. We tried $|d_{(j)}|^{2}$ and $|d_{(j)}|$ and did not gain any improvement in the results. 

We use autoencoders to reduce the dimension of $S(d_{(j)})$ in order to apply the calibration procedure. The autoencoder is trained with an Adam optimizer \citep{Kingma:2014} and a learning rate, $10^{-4}$, is chosen by minimizing the mean squared error. The size of $S(d_{(j)})$ is then reduced from $2^{16}$ to $2^3$, which is far less than the size of the overall training data set. From our preliminary study, reducing the size of the data set any less than $2^{3}$ gave a significantly poor fit. For classification, an artificial neural network (ANN) with one fully-connect layer is trained using the cross entropy loss function. The calibration curve is obtained from the feed-forward ANN inversion. We tried multiple layers for the ANN but found no real gain from using more than one layer for this toy example. 

\begin{figure*} 
\begin{minipage}{0.5\textwidth}
    \centering
    \includegraphics[width=\linewidth]{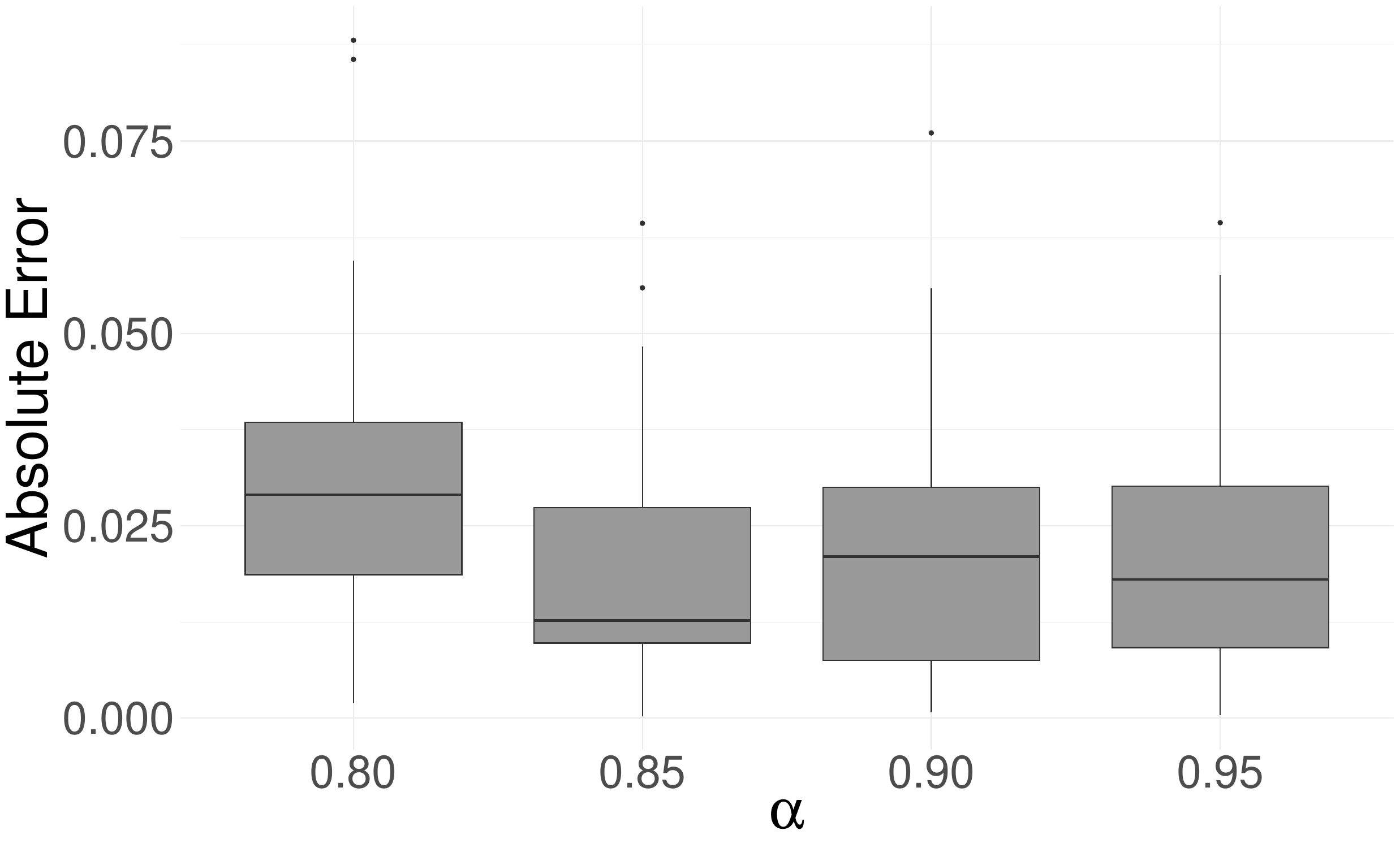}
  \end{minipage}%
  \begin{minipage}{0.5\textwidth}
    \centering
    \includegraphics[width=\linewidth]{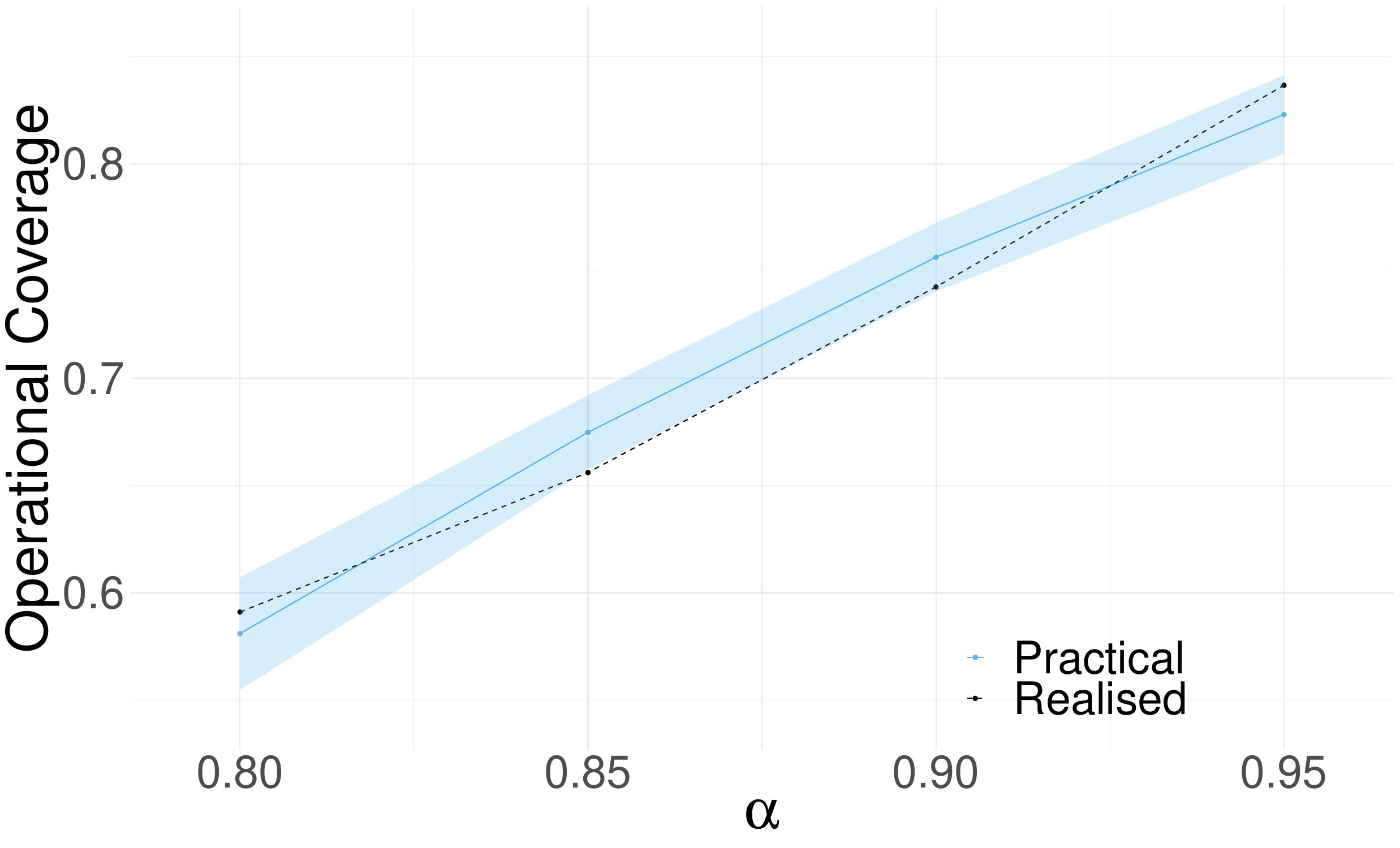}
  \end{minipage}

\caption{Estimation of $\dot{f}$-only. Absolute error from 30 replicates against $\alpha$ (left) and operational coverage estimates $\bar{b}_r(S(d_o))$ with 2-SE error and {\it realised operational coverage} $b_r(d_o)$ (right) for the test waveform $d_o$.} 
\label{fig:sim1D} \end{figure*}

The left panel of Figure \ref{fig:sim1D}  presents the performance of the operational coverage estimator (with summary of the procedure given in Figure \ref{fig:process}) using 30 approximate signals with noise and on average absolute errors are relatively small. The right panel of Figure \ref{fig:sim1D} compares the operational coverage estimates using the {\it practical estimator} $\bar{b}_r(S(d_o))$ (Section \ref{subsec:practicaloperationalcoverage} and the {\it realised operational coverage} $b_r(d_o)$ (\ref{eq:realised}) for the test signal. At a given nominal level $\alpha$, the practical operational coverage and the realised operational coverage agree to excellent precision. Absolute errors tend to be less than 0.06 in general and, for the test data, the coverage of posterior density over the 0.9 approximate credible set is 0.76.

\begin{figure*} 
\begin{minipage}{0.5\textwidth}
    \centering
    \includegraphics[width=\linewidth]{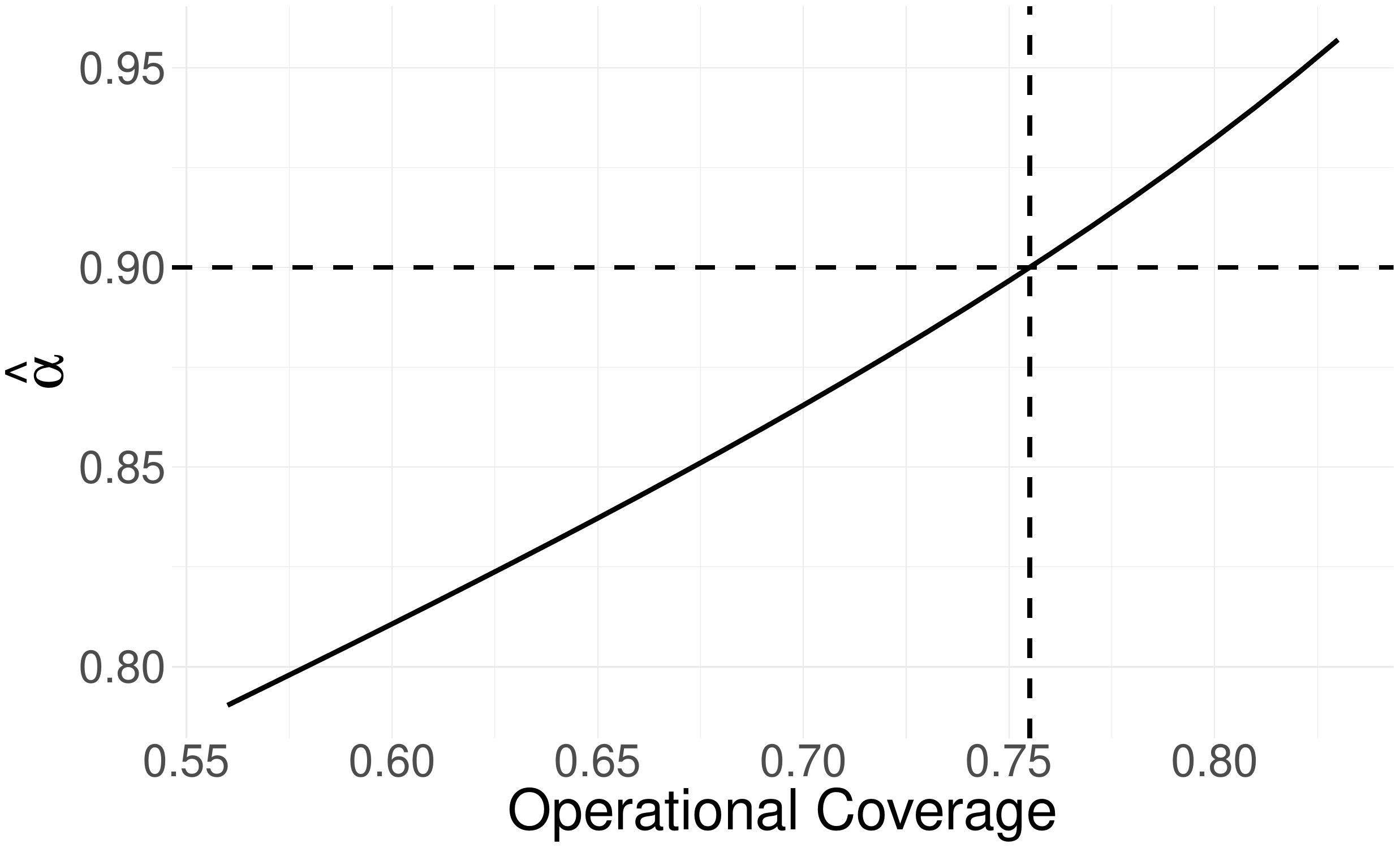}
  \end{minipage}%
  \begin{minipage}{0.5\textwidth}
    \centering
    \includegraphics[width=\linewidth]{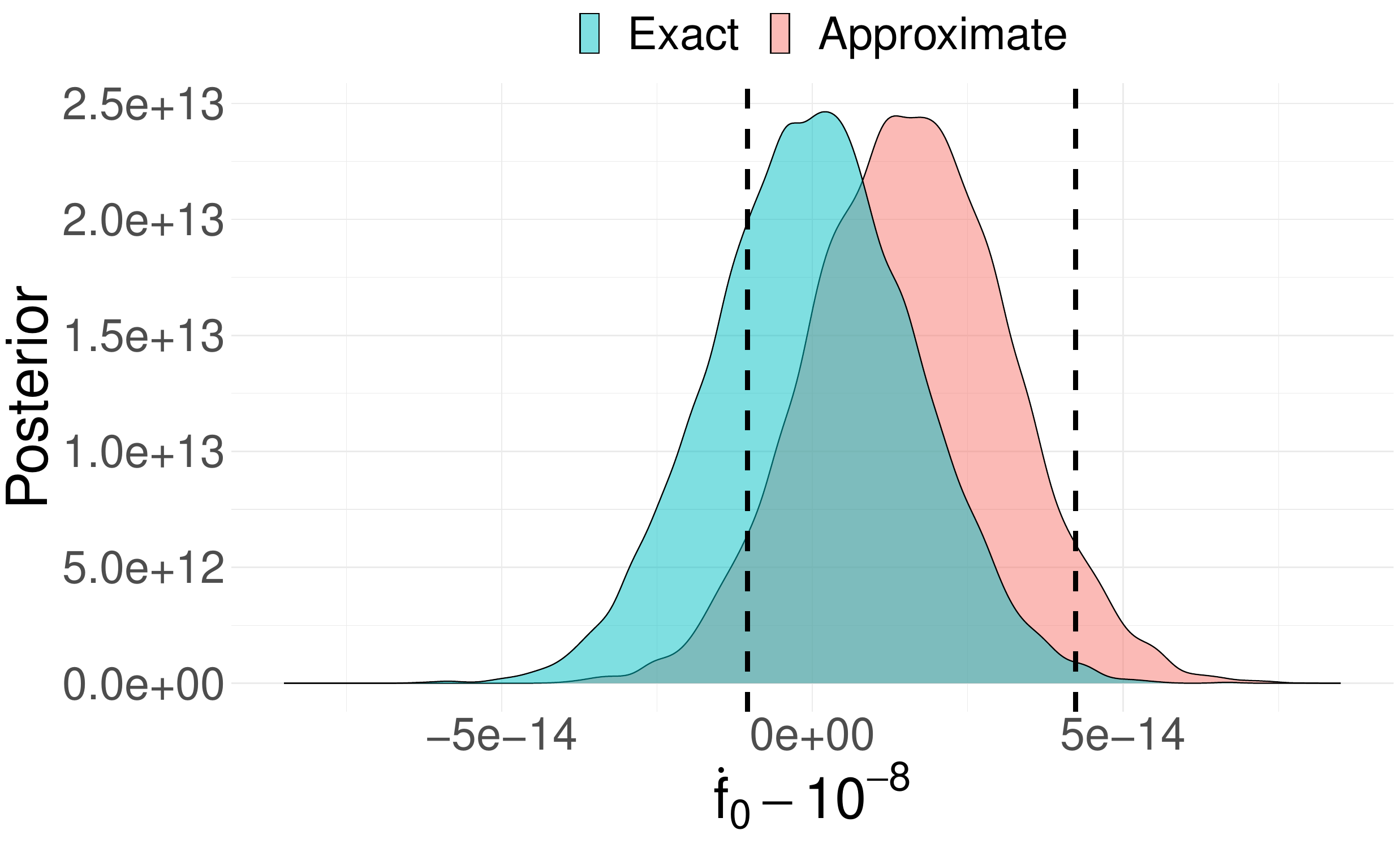}
  \end{minipage}
\caption{Left panel: Calibration curve. Right panel: Exact and approximate posterior densities of $\dot{f}_0$. The dashed lines on the top figure represent the operational coverage for the 90\% credible interval of approximate posterior ($\widetilde{C}_{d_o,0.9}$). This is calculated by using the calibration curve to map the calibrated nominated level of $\hat{\alpha} = 90\%$ back to an operational coverage level, which is $\approx 76\%$ for the test waveform.}  \label{fig:test1D} \end{figure*}

Exact and approximate inferences on the true signal \eqref{eq:exact_toy_model} are compared in Figure \ref{fig:test1D}. It is observed that the exact and approximate posterior densities do not overlap completely and some deviation between them. The calibration curve shows how the calibrated level changes with the desired operational coverage and the calibrated level $\widehat{\alpha}$ is higher than $b(d_o)$. i.e., $\bar{b}_r(S(d_o))$ is smaller than $\alpha$ (right panel of Figure \ref{fig:sim1D}). The main conclusion of this study is that we have calibrated an approximate credible set (arising from an approximate posterior) to achieve 0.76 coverage of the posterior.

We now consider the full model with three parameters. i.e., $\boldsymbol{\theta} = \{a_{0}, f_{0}, \dot{f}_{0}\}$. Tight priors are assigned on these three parameters 
\begin{align*}
    a &\sim \text{U}[a_{0}-10^{-22}, a_{0}+10^{-22}]\\
    f &\sim \text{U}[f_{0}-10^{-7}, f_{0}+10^{-7}]\,\text{Hz}\\
    \dot{f} &\sim \text{U}[\dot{f}_{0}-10^{-13}, \dot{f}_{0}+10^{-13}]\, \text{Hz}/\text{s}. 
\end{align*}

We increase the size of the training data to 10,000, $\{S(d_{(j)}), c_{j}, \alpha_{j}\}_{j=1}^{10000}$ where $S(d_{(j)})=\text{Re}(d_{(j)})$ and it is generated similarly. For approximate posterior density, the multivariate Gaussian form of density approximation for logged parameter values was imposed. We used three fully connected layers with one dropout layer in both the encoder and decoder to reduce the dimension of $S(d_{(j)})$ from $2^{16}$ to $2^6$. A 1-layer ANN classifier with the $l_1$ penalty on weights of neurons is fitted. 

\begin{figure*}
\begin{minipage}{0.5\textwidth}
    \centering
    \includegraphics[width=\linewidth]{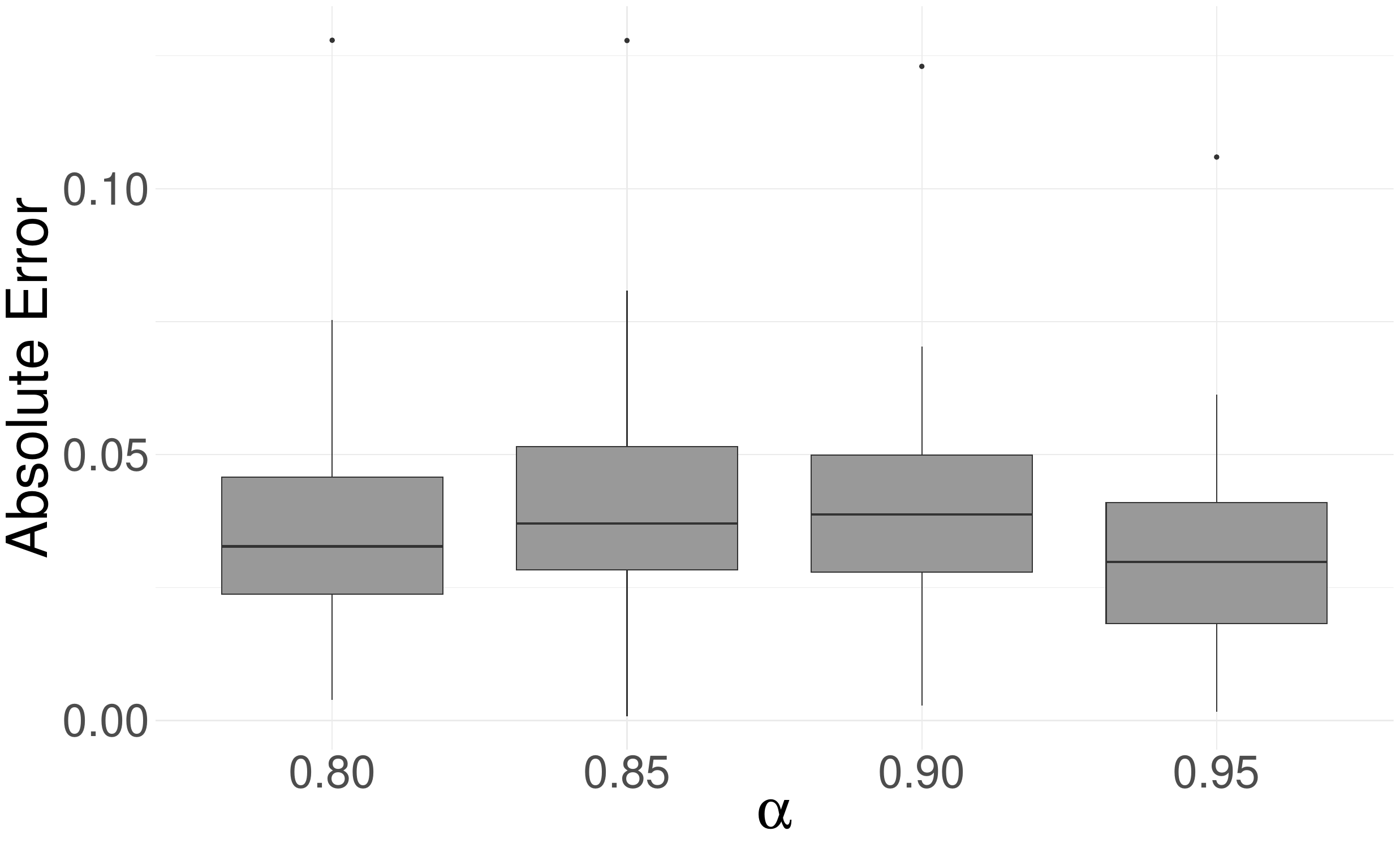}
  \end{minipage}%
  \begin{minipage}{0.5\textwidth}
    \centering
    \includegraphics[width=\linewidth]{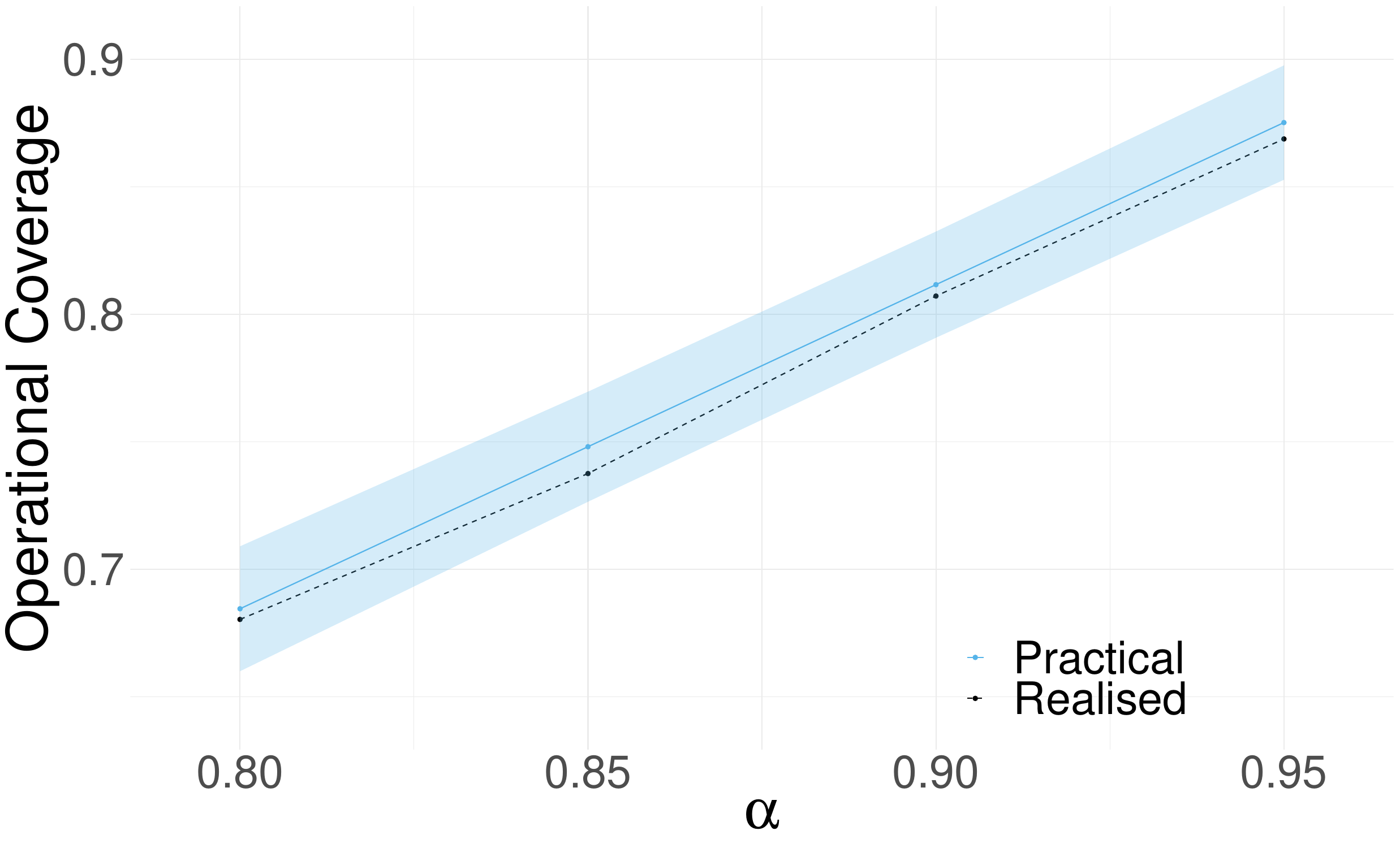}
  \end{minipage}
\caption{Estimation of the full model, i.e. $a_{0}, f_{0}, \dot{f}_{0}$. Absolute error from 30 replicates against $\alpha$ (left) and operational coverage estimates $\bar{b}_r(S(d_o))$ with 2-SE error and {\it realised operational coverage} $b_r(d_o)$ (right) for the test waveform $d_o$.} \label{fig:sim3D} \end{figure*}

\begin{figure*}
\includegraphics[width=0.8\linewidth]{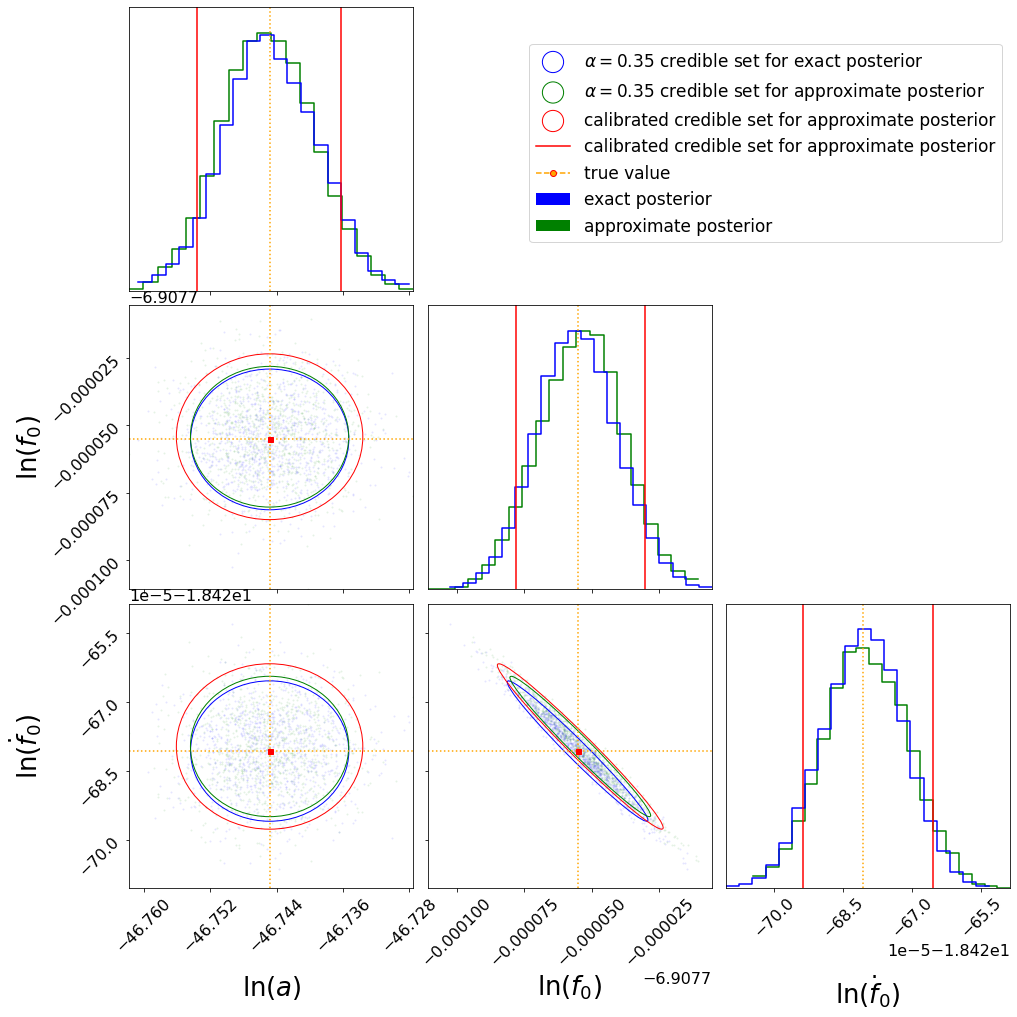} 
\caption{Corner plot summary of samples from the exact and approximate posterior densities in the toy model}. The panels on the diagonal show the exact (blue) and approximate (green) empirical marginal posterior densities with the true value (yellow dashed line) and calibrated approximate credible interval (red line). The off-diagonal panels shows the exact (blue contour) and approximate (green contour) credible sets with a nominal level of $\alpha=0.8$ and, calibrated approximate credible set (red contour) with the desired operational coverage of 0.8 based on exact (blue points) and approximate (green points) posterior sample. The true values $\boldsymbol{\theta}_0$ are marked by red points. 
 \label{fig:test3D} \end{figure*}

\begin{figure}
\includegraphics[width = 1\linewidth]{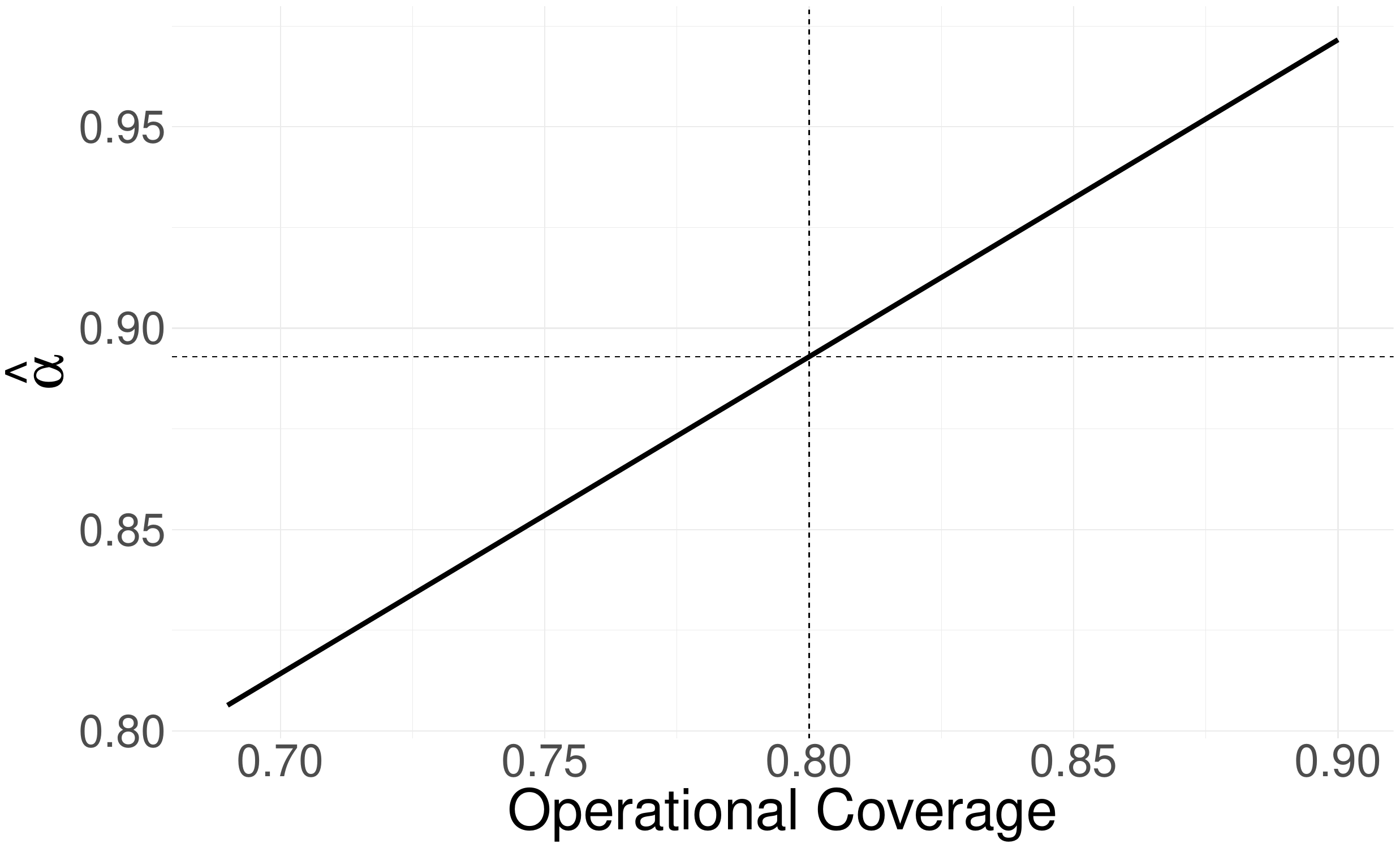}
\caption{Calibration curve for $d_o$. The dashed lines represent the calibrated nominal level $\widehat{\alpha}=0.89$ for the target operational coverage $b(d_o)=0.8$. }\label{fig:test3D_rmse} \end{figure}

The performance of the operational coverage estimator using 30 approximate signals with noise is shown in the left penal of Figure \ref{fig:sim3D} and absolute errors are less than $\approx$0.075 in general. The right penal of Figure \ref{fig:sim3D} compares $\bar{b}_r(S(d_o))$ and $b_r(d_o)$ for the test signal. The practical and realised operational coverage agree relatively well at a given $\alpha$. Exact and approximate posterior densities for the test data $d_o$ are compared in Figure \ref{fig:test3D}. Calibration curve for the test waveform $d_o$ is shown in Figure \ref{fig:test3D_rmse} and the calibrated nominal level $\widehat{\alpha}$ is higher than the desired operational coverage value, i.e., $b_r(d_o)$ is smaller than $\alpha$. For $d_o$, the coverage of posterior density over the approximate credible set with a nominal level of 0.89 is 0.8. 

Having demonstrated our calibration procedure through an illustrative example, we will now apply it to a more realistic gravitational-wave scenario.

%%%%%%%%%%%%%%%%%%%%%%%%%%%%%%%%%%%%%%%%%%%%%%%%%%%%%%%%%%%%%%%%%%%%%%%%%%%%%%%%%%%%%%%
\section{Application: Massive black holes}\label{sec:massive_black_holes_application}

In this section, we apply the calibration procedure discussed in section \ref{sec:general_calibration_methodology} and exemplified in section \ref{sec:toy_model} on a realistic massive black hole binary signal. Using \texttt{lisabeta} developed in~\cite{marsat2021exploring, marsat2018fourier}, we generate complete inspiral-merger-ringdown frequency domain spin-aligned massive black holes in the solar system barycenter frame with the LISA response applied. 

\subsection{Set up}
In the Bayesian inference, we incorporate higher modes $\mathcal{H}_{e} = \{(2,2),(2,1),(3,3),(3,2),(4,4),(4,3)\}$ for the true MBH waveform denoted by $h_e(t;\boldsymbol{\theta})$. Approximate waveforms denoted by $h_m(t;\boldsymbol{\theta})$ are generated by removing a single harmonic $\mathcal{H}_{m} = \mathcal{H}_{e} \backslash \{(4,3)\}$, giving a deviation from the exact model and the approximate model. For our exact signal with modes $\mathcal{H}_{e}$, we define the true parameters: the total mass $M = m_{1} + m_{2} = 3\cdot 10^{7}M_{\odot}$; mass ratio $q = m_{1}/m_{2} = 2$; the two effective spin parameters of the two component masses $\chi_{1} = 0.5$ and $\chi_{2} = 0.5$; the time of coalescence $t_{c} = 10^{5}\,$seconds; luminosity distance $6.67\,$ Gpc; initial phase at coalescence $\phi_{c} = 1.1$; sky position $(\beta = 0.3, \lambda = 0.8)$ in ecliptic coordinates; and polarisation angle $\psi = 1.7$. In our example, we will focus on a subset of parameters $\boldsymbol{\theta} = \{M, q, \chi_{1},\chi_{2},t_{c}\}$, to demonstrate the calibration procedure as a proof of principle.

We choose true parameters given by $\boldsymbol{\theta}_0 = \{M_{0} = 10^{7}M_{\odot}, q_{0} = 2, \chi_{1,0} = 0.5, \chi_{2,0} = 0.5, t_{c,0} = 10^{5}\,\text{seconds}\}$. The observation time of our signals will be $\sim 1$ day, sampled with cadence $\Delta t = 200$ seconds. Due to the large total mass, the frequencies emitted by the massive black hole signal are low, even at the larger harmonics. This allows us to analyse very short data segments with large sampling intervals thus reducing computational costs. The length of our data sets are $N = 2^{12}$. We found that the optimal matched filtering SNR using \eqref{eq:SNR_X} over both the A and E channels are given by $\rho_{A} \sim 2022.02$ and $\rho_{E}\sim 1702.57$ giving a total SNR over both A and E as $\rho_{AE} \sim 2643.35$.

Before applying the calibration procedure, we will show that our Fisher matrix calculations are not subject to numerical instabilities. We inject a signal with true parameters $\boldsymbol{\theta}_0$ defined earlier into a two noiseless data streams corresponding to TDI channels $X = \{A, E\}$. Including the $T$ channel is unnecessary for our purposes since the contribution of SNR is low with respect to the $A$ and $E$ channels. For inference, we use the likelihood defined in \eqref{eq:whittle_likelihood_signal} with model template $h_{m}$ with an incomplete set of modes $\mathcal{H}_{m} = \mathcal{H}_{e}/\{(4,3)\}$. Starting close to the true parameters, we use the \texttt{emcee} sampling algorithm to generate samples from the approximate posterior density $\tilde{p}(\boldsymbol{\theta}|d)$. To compute the Fisher matrix, we use the numerical procedure of finite differences to compute derivatives of the MBH waveform with respect to parameters. After computing the matrix \eqref{eq:FM_AET}, we apply a log-transformation to reduce the condition number of the matrix prior to computing the inverse. From equation \eqref{eq:bias_mismodelling_expectation}, we then sample from a multivariate Gaussian

\begin{align}\label{eq:sample_multivariate_gaussian_FM_AET}
\begin{split}
\boldsymbol{\theta} &\sim \mathcal{N}\left(\boldsymbol{\theta}_{0} + \boldsymbol{\theta}_{\text{bias}}, \Gamma_{AE}^{-1}(h_{m})\right)\, , \\
\theta^{i}_{\text{bias}} &= [\Gamma_{AE}(h_{m})^{-1}]^{ij}\sum_{X = \{A,E\}}(\partial_{j}h^{(X)}_{\text{m}}|\delta h^{(X)} + n^{(X)} )
\end{split}
\end{align}
for $\theta^{i}_{\text{bias}}$ a component of $\boldsymbol{\theta}$. We remind the reader that each of the quantities in \eqref{eq:sample_multivariate_gaussian_FM_AET} is evaluated at the true parameters $\boldsymbol{\theta}$. We then plot the histogram of samples alongside the approximate posterior density in Figure \ref{fig:5dFMcheck}. What we learn here is that the Fisher matrix is a suitable approximation to the posterior density and can be used to approximate posterior distributions generated using an approximate waveform model. 

\subsection{Calibration Procedure}
As a proof of principle, we will apply our calibration procedure on a five-dimensional space. We will choose the five parameter set  $\boldsymbol{\theta} = \{M,q,\chi_{1},\chi_{2},t_{c}\}$. For this study, tight uniform priors are set for the five parameters as follows:
\begin{eqnarray*}
    M &\sim&  \text{U}[ M_{0} -5\times 10^4, M_{0} + 5\times 10^4]\\
    q &\sim& \text{U}[q_{0} -2.5\times 10^{-3}, q_{0} + 2.5\times 10^{-3}]\\
    \chi_1 &\sim&  \text{U}[\chi_{1,0} -5\times 10^{-4},\chi_{1,0}+ 5\times 10^{-4}]\\
    \chi_2 &\sim&  \text{U}[\chi_{2,0} - 5\times 10^{-4},\chi_{2,0} + 5\times 10^{-4}]\\
    t_{c} &\sim&  \text{U}[t_{c,0}-0.25,t_{c,0} + 0.25].
\end{eqnarray*}

Following a similar procedure outlined in \ref{sec:calibration_procedure_toy}, the data stream is $d=\hat{h}_e(\boldsymbol{\theta})+\hat{n}$ and the test waveform is $d_o=\hat{h}_e(\boldsymbol{\theta}_0)$. The training data $\{S(d_{(j)}),c_j,\alpha_j \}^{10^5}_{j=1}$ where $S(d_{(j)})=\text{Re}(d_{(j)})$ is generated to get a practical operational coverage estimator. We also considered $|d_{(j)}|^{2}$ and did not gain any significant improvements in the results and, the corresponding result is not included in the paper. We reduce the input feature size of $N = 2^{12}$ to $N = 2^{8}$ using a three-layer fully-connected encoder and decoder network. For classification, an ANN using three fully-connected hidden layers and an output layer with a sigmoid activation function are used. For the {\it realized operational coverage} estimation $b_r$, we used 26,880 exact posterior samples, using the complete IMR waveform with full harmonic structure $\mathcal{H}_{e}$, with 32 parallel chains with a thinning factor of 5. 

\begin{figure*}[b]
\includegraphics[width = 0.8\linewidth]{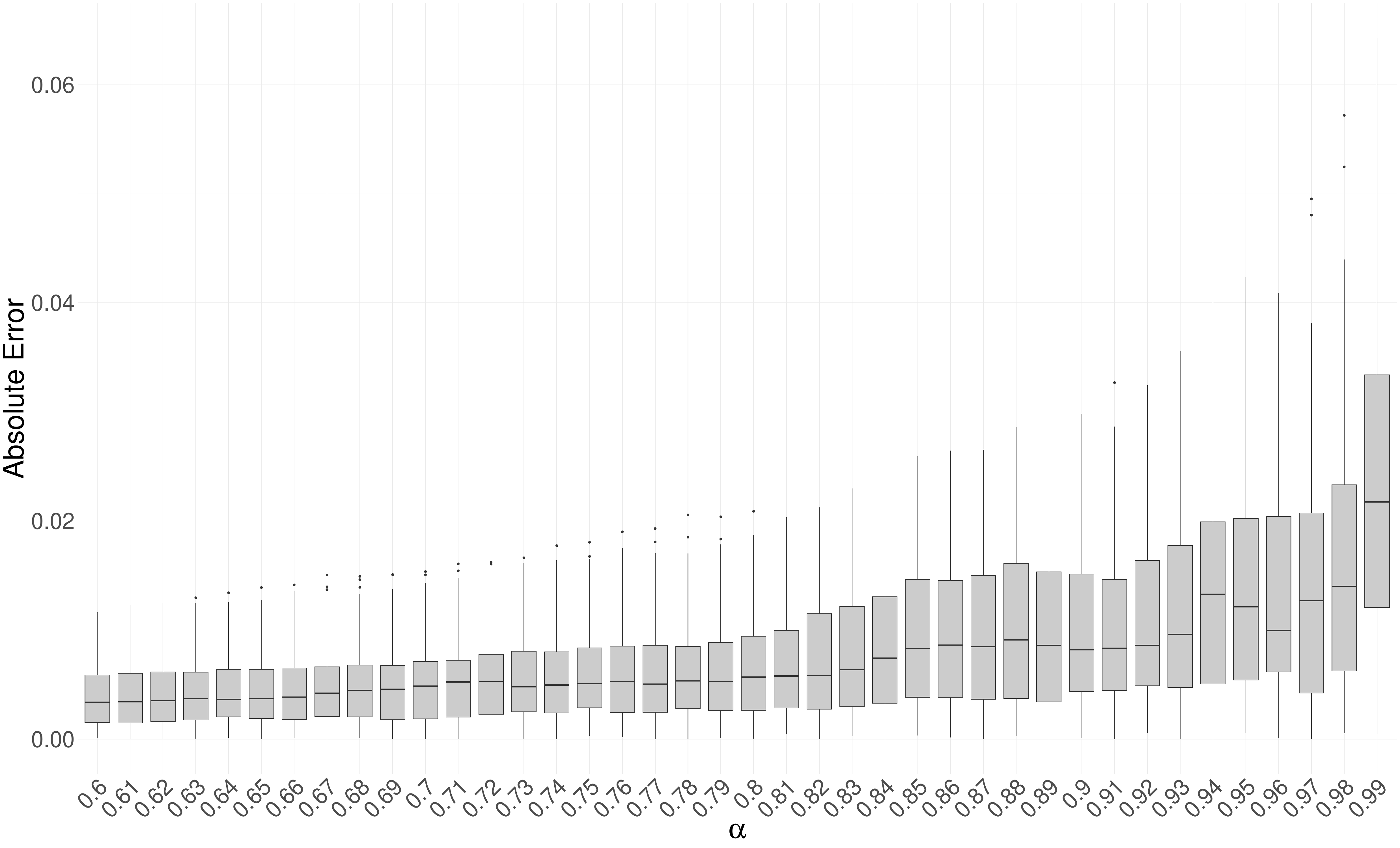} 
\caption{Absolute error between operational coverage estimate and realised operational coverage against $\alpha$ from 30 replicates. The black dots are outlier absolute errors at a specific nominal level $\alpha$. 
} \label{fig:test5D_rmse} \end{figure*}

\begin{figure*} 
\begin{minipage}{0.5\textwidth}
    \centering
    \includegraphics[width=\linewidth]{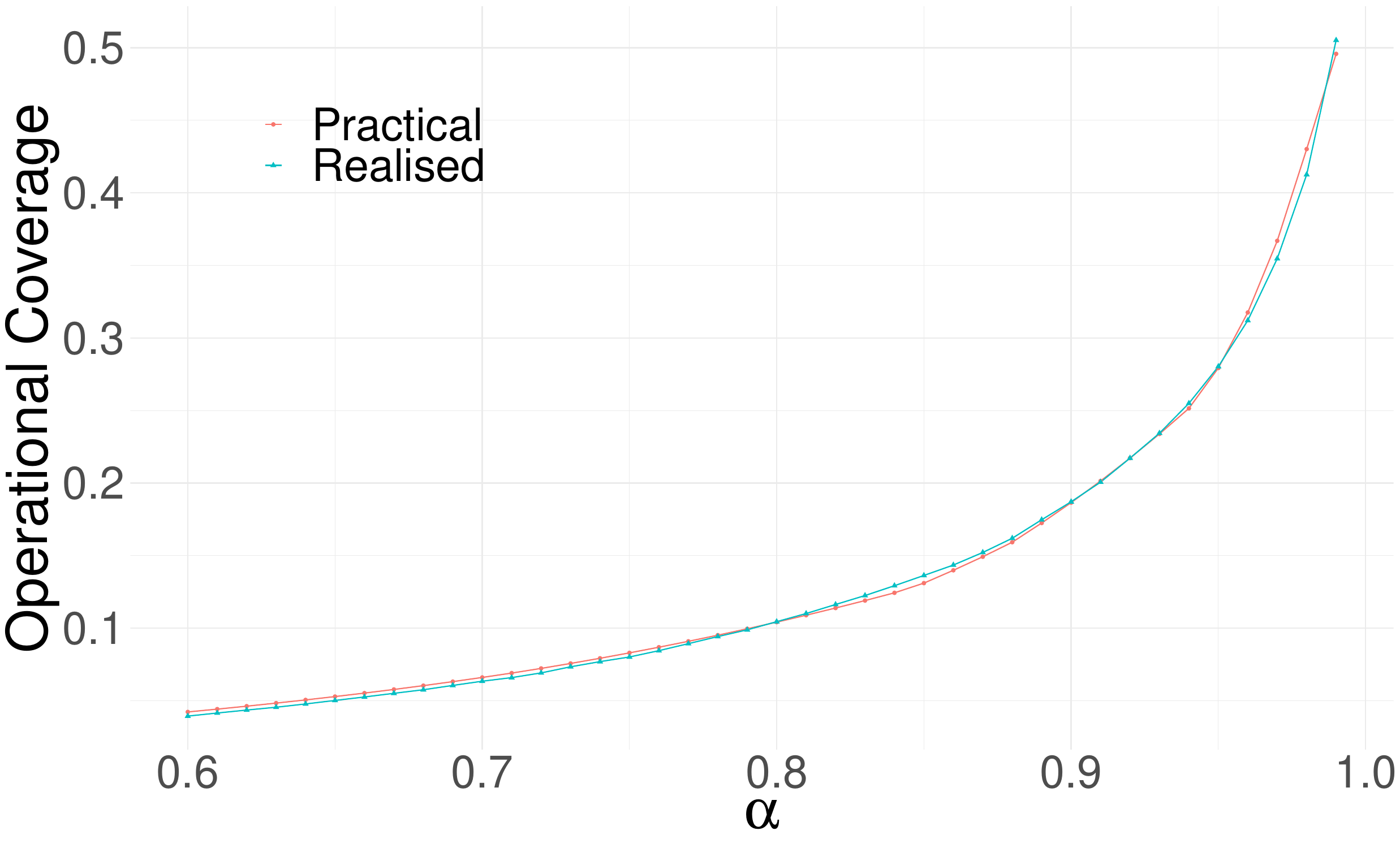}
  \end{minipage}%
  \begin{minipage}{0.5\textwidth}
    \centering
    \includegraphics[width=\linewidth]{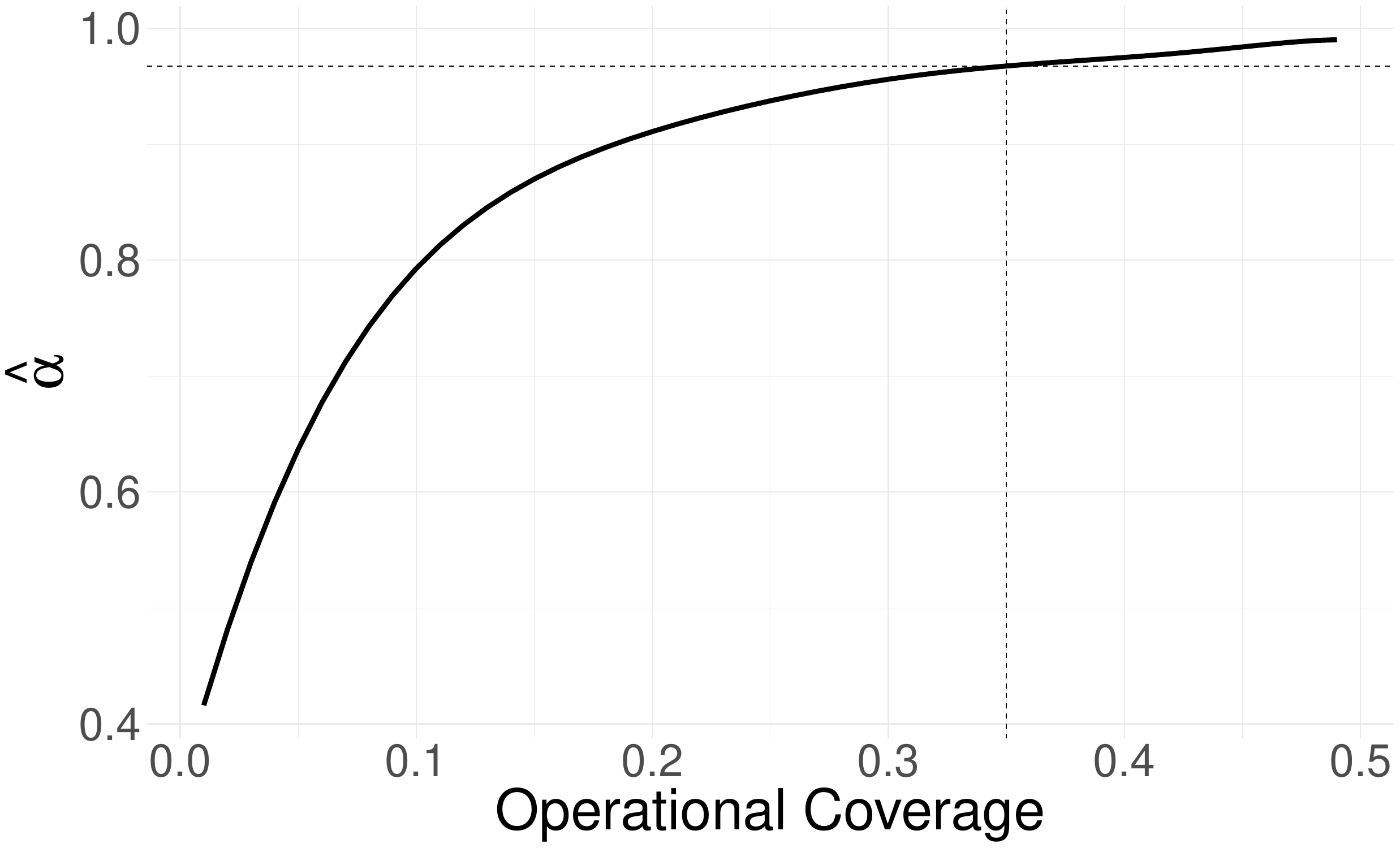}
  \end{minipage}
\caption{Top panel: Operational coverage estimates $\bar{b}_r(S(d_o))$ (red) and realised operational coverage $b_r(d_o)$ (blue) for the test data $d_o$ against $\alpha$. Bottom panel: Calibration curve for $d_o$ (Calibrated nominal levels $\widehat{\alpha}$ against the target operational coverage $b(d_o)$). The dashed lines represent the calibrated nominal level $\widehat{\alpha}=0.97$ for the target operational coverage $b(d_o)=0.35$. }\label{fig:GWt5D_map} 
\end{figure*}

\begin{figure*} [b]
\includegraphics[width=16cm,height=16cm]{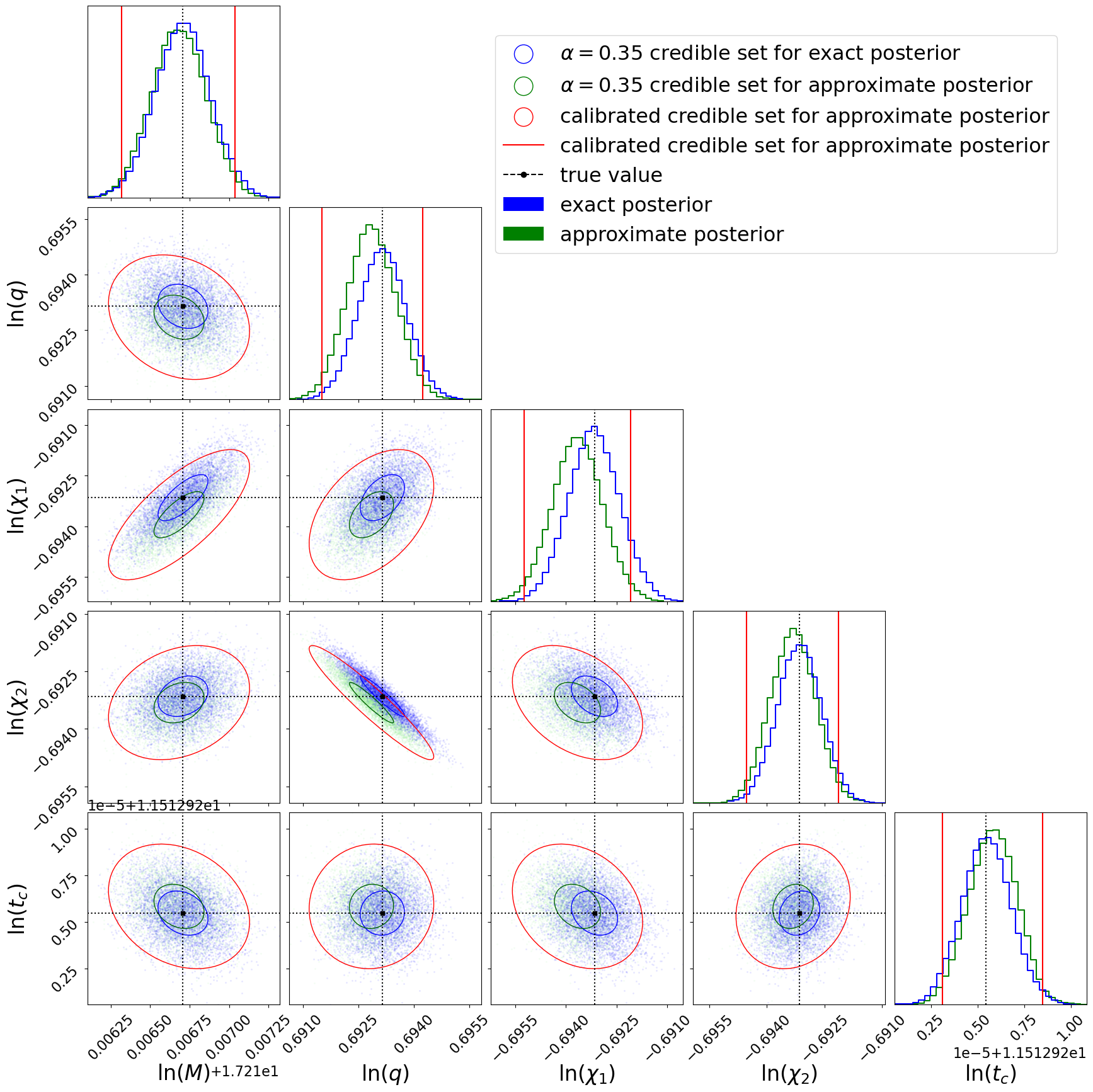} 
\caption{
Corner plot summary of samples from the exact and approximate posterior densities for the massive black hole binary signal $d_o$. The panels on the diagonal show the exact (blue) and approximate (green) empirical marginal posterior densities with the true value (black dashed line) and calibrated approximate credible interval (red line). The off-diagonal panels show the exact (blue contour) and approximate (green contour) credible set with a nominal level of $\alpha=0.35$ and, the calibrated approximate credible set (red contour) with the desired operational coverage of $0.35$ based on exact (blue points) and approximate (green points) posterior samples. The true values $\boldsymbol{\theta}_0$ are marked by black points.
} \label{fig:GW5D} \end{figure*}

In general, the operational coverage estimator $\bar{b}_r$ exhibits generally small absolute errors in Figure \ref{fig:test5D_rmse}. Although the absolute errors tend to increase with $\alpha$, relative errors are likely to be less variable. The absolute error is small as 0.013 and large as 0.065. i.e., the smallest and largest relative absolute errors are $0.013/0.6=0.0217$ and $0.065/0.99=0.0657$ respectively. For the test waveform $d_o$, the estimate $\bar{b}_r(S(d_o))$ from the practical operational estimator is compared to the realistic operational coverage $b_r(d_o)$ in the top plot of Figure \ref{fig:GWt5D_map}. The calibration curve in Figure \ref{fig:GWt5D_map} was modelled by a polynomial regression with the degree 7. We observe a very small operational coverage in comparison to the nominal level ($ \bar{b}_r(S(d_o)) \ll \alpha$), and this is due to the small overlap between the exact and approximate posterior distributions in Figure \ref{fig:GW5D}, in particular $q$ and $\chi_2$.  For example, the posterior coverage of the 0.967 credible set of an approximate posterior distribution is 0.35 and this is also confirmed from the calibration curve that the calibrated level is 0.97 for the target operational coverage of 0.35. As result, the calibrated credible set of an approximate posterior (red contour) is larger in order to achieve the target 0.35 operational coverage, which is about 1.8$\sigma$.

%\onecolumngrid

%%%%%%%%%%%%%%%%%%%%%%%%%%%%%%%%%%%%%%%%%%%%%%%%%%%%%%%%%%%%%%%%%%%%%%%%%%%%%%%%%%%%%%%%%%%%%

%\twocolumngrid
\section{Discussion and conclusions}\label{sec:discussions_conclusions}

In this paper, we have presented a Bayesian calibration procedure for the operational coverage of an approximate family of GW posteriors and applied the method to both a toy model and a more realistic analysis of MBHB signals for LISA. Specifically, the posteriors that are calibrated in our work are Fisher-information-based approximations (i.e., normal approximations) to the posteriors that are obtained by using an approximate waveform model in the GW likelihood. Such approximations are frequently used in GW astronomy to perform bulk calculations in exploratory data analysis studies; our proposed method then allows users to rapidly obtain corrected credible sets that correspond to some desired coverage level (``correct'' relative to a more accurate waveform model).

At present, our proposed method is novel in the GW literature and has no comparable counterpart since other methods for inference calibration i) focus on correcting the posterior itself in terms of coverage and ii) deal only with a single posterior at a time. For example, one would have to evaluate the accurate model in bulk in order to use the method proposed by Cutler and Vallisneri \cite{Cutler:2007mi} for estimating inference biases on the fly. Even if a regression model is fit directly to the biases over the waveform model space, that would still require learning a parameter vector rather than a single coverage number, which would require more complex computation and stringent accuracy requirements on the fit.

In addition to our proposed usage of the method in bulk studies over a space of GW signals, the calibration procedure could potentially also be used on actual data containing a signal. With a model that is pre-trained on more accurate waveforms and realistic detector noise, one could rapidly compute calibrated credible sets for the approximate waveforms used in template banks for ground-based observing. This could be useful for applications such as the rapid sky localisation of sources for low-latency electromagnetic follow-up. The future third-generation ground-based GW detectors, such as the Einstein Telescope and Cosmic Explorer, will have enhanced sensitivity to gravitational wave signals in the Hz frequency band $f\in (5, 2000)\,\text{Hz}$. These instruments can exploit the proposed operational coverage discussed in this paper in order to quantify the systematic biases and thus investigate the impact of inaccurate waveforms on tests of GR in an efficient way~\cite{PhysRevResearch.2.023151}. The estimated operational coverage can be regarded as a criterion to set requirements for the sensitivity of the detectors to yield unbiased parameter inference for the target system. The forward-thinking calibrated result gives reasonable and meaningful information for future GW research. 

The MBH example we presented was restricted to five parameters of the full waveform model. However, the computational complexity of the calibration procedure --- in particular, the operational coverage estimator --- will typically require a significant increase in computing resources as additional parameters are considered. Increasing the computational efficiency of the operational coverage estimator for high-dimensional problems, and thus improving the scalability of the calibration method to the dimensionalities of both the parameter space and the data representation, is an avenue for future research. 

We should point out that the approximate posterior distribution is not too different to the true one. At least, the overlap of the credible intervals exists, which is not always the case. As the divergence between posteriors becomes large, the nominal credible level to achieve any operational coverage goes to 1, which is trivial and manifests the inability to train the model. Therefore, it is encouraged to estimate the maximum operational coverage when obtaining the calibration curve; if it is  small, for example, less than $0.5\sigma$, more accurate waveforms should be considered. 

The \texttt{Python} code for the one-dimensional toy example in Section \ref{sec:toy_model} is provided at \url{https://github.com/bpandamao/calibration_case_study}. 

\section*{Acknowledgements}

JEL, MCE and R.~Meyer acknowledge support by the
Marsden grant MFP-UOA2131 from New Zealand Government funding, administered
by the Royal Society Te Aparangi. 
AJKC acknowledges previous support from the NASA LISA Preparatory Science grant 20-LPS20-0005. OB acknowledges support from the French space agency CNES in the framework of LISA. He also thanks Sylvain Marsat for giving permissions to use \texttt{lisa-beta}. All computations are performed on a virtual machine with 32GB RAM, 16 VCPUs, and an Ubuntu Linux operating system. The autoencoder and ANN were implemented using Python package \texttt{Tensorflow} and \texttt{Scikit-Learn}. We thank the Center for eResearch (CeR) at the University of Auckland for providing access to and assistance with  the Nectar Research Cloud. 

\section*{Author Contributions}

R.~Mao: Conceptualization, Data curation, Formal analysis, Investigation, Methodology, Project administration, Resources, Software, Supervision, Validation, Visualization, Writing – original draft. 

JEL: Conceptualization, Data curation, Investigation, Methodology, Project administration, Resources, Software, Supervision, Validation, Visualization, Writing – original draft. 

OB: Conceptualization, Data curation, Formal analysis, Investigation, Software: Fisher matrix, MCMC \& \texttt{lisa-beta}, Methodology, Resources, Software, Supervision, Validation, Visualization, Writing – original draft. 

AJKC: Conceptualization, Formal analysis, Methodology, Project administration, Supervision, Writing – original draft, Writing - Review \& Editing.

MCE: Conceptualization, Methodology, Project Administration, Software, Visualization, Writing - Original Draft, Writing - Review \& Editing.

R.~Meyer: Conceptualization, Methodology, Funding acquisition, Project Administration, Supervision, Writing - Original Draft, Writing -- Review \& Editing.

\clearpage % Start a new page

\appendix
\onecolumngrid
\section{Fisher matrix computations}

\begin{figure*}[b]
\includegraphics[width=14cm,height=14cm]{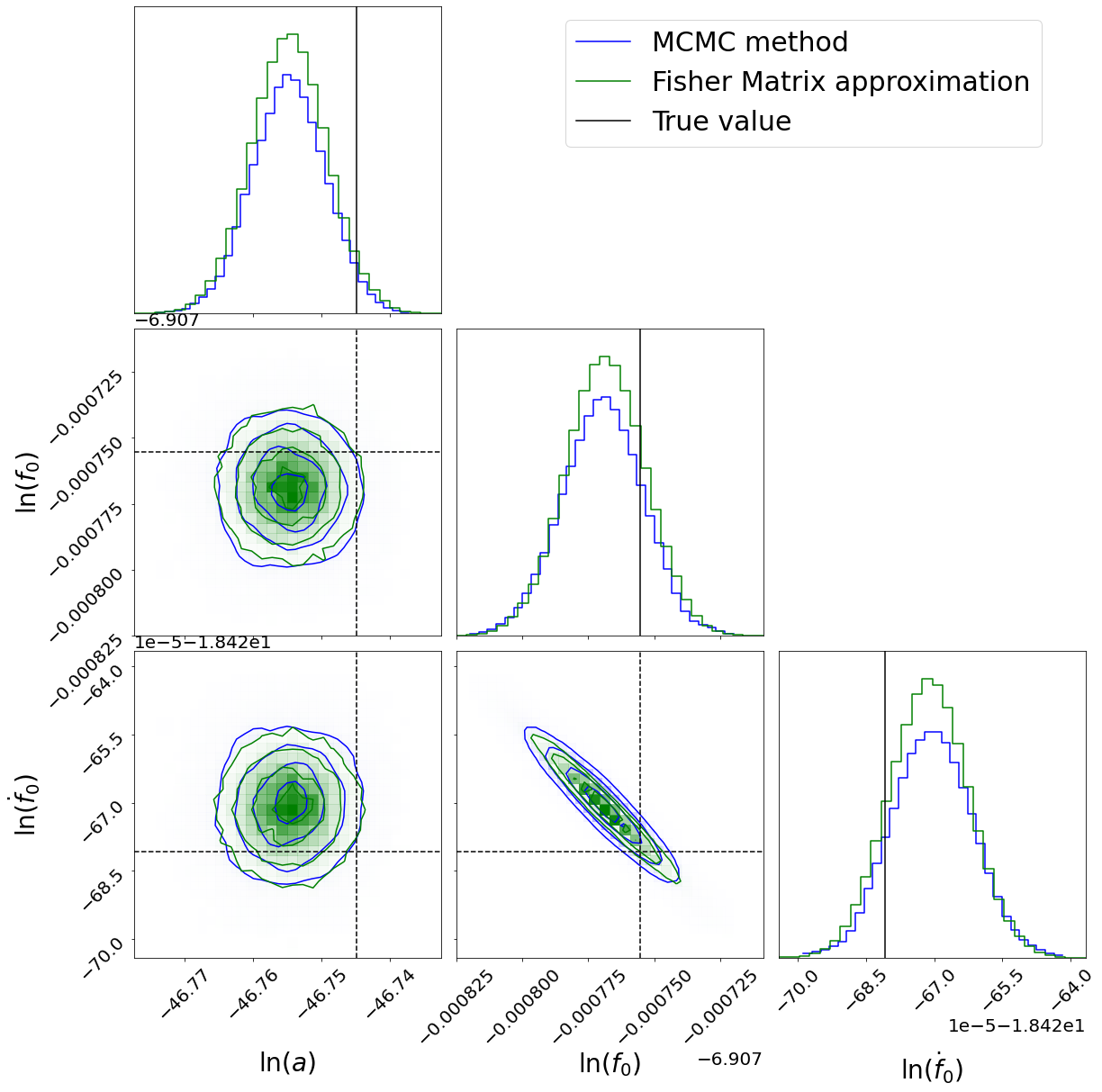}
\caption{ The blue curve represents a posterior distribution on parameters $\boldsymbol{\theta}$ when inferring parameters of an injected exact signal $h_{e}(t;\boldsymbol{\theta}_{0})$ into Gaussian noise with an approximate model template $h_{m}(t;\boldsymbol{\theta},\epsilon = 10^{-6})$. The green curve represents an approximation to the posterior, computed via the Fisher matrix formalism. We highlight here that the computation of the Fisher matrix accurately describes the posterior, implying we are not subject to numerical instabilities when calculating derivatives/inverses. The black line indicates the value of the true parameters in the study.} 
\label{transformationg:3dFMcheck} 
\end{figure*}

\begin{figure}
\includegraphics[width=15cm,height=15cm]{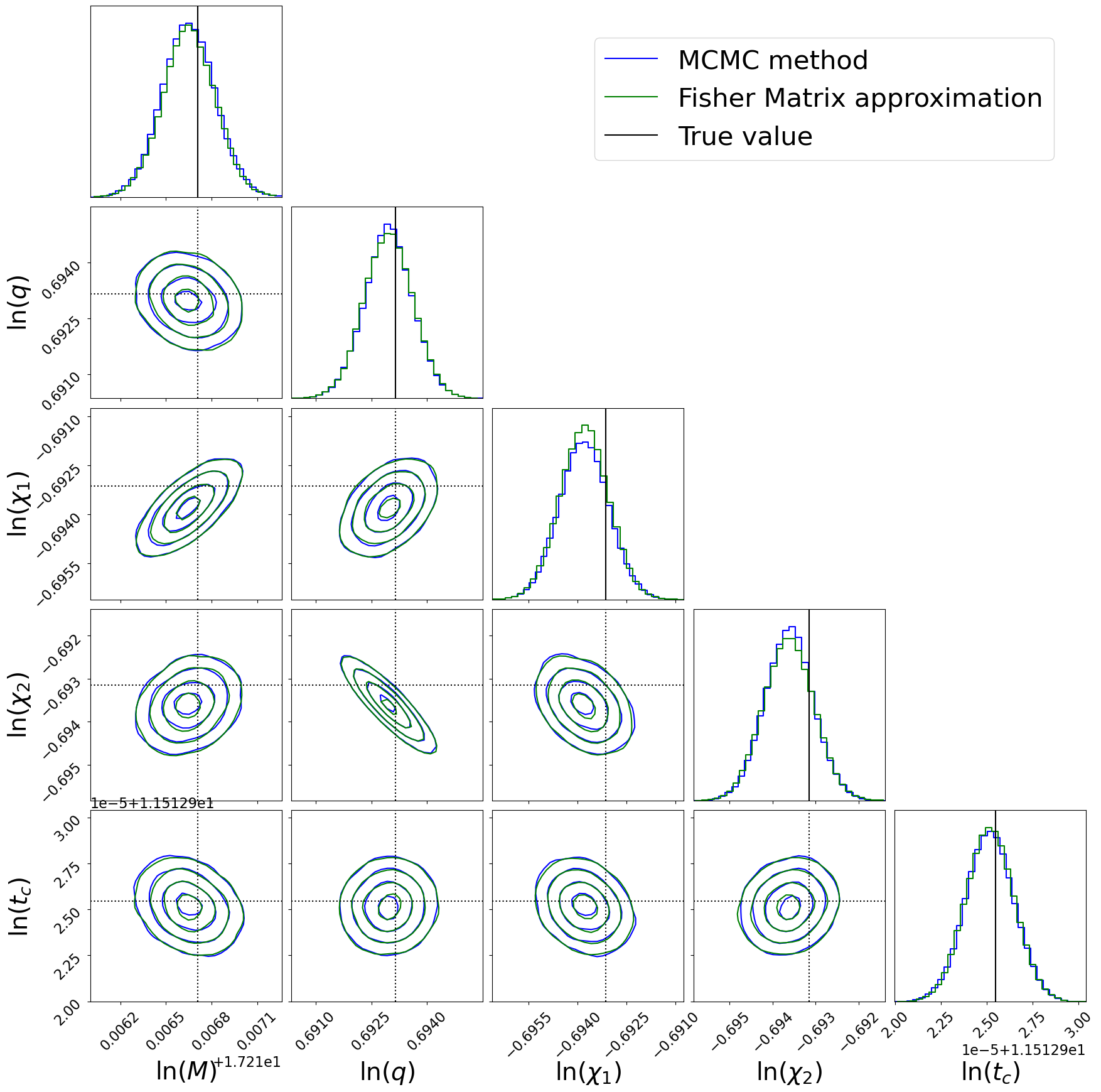} 
\caption{Comparison of posterior distribution between the MCMC method (blue) and the Fisher Matrix approximation (green) at the massive black hole binary signal $d_{o}$, with black vertical/horizontal lines denoting the true parameters. We highlight here an excellent agreement between our MCMC simulation and the approximate Fisher matrix approach. We see that both the fluctuations to recovered parameters (induced through noise via equation \eqref{eq:noise_fluc} and precision measurements on parameters encoded by \eqref{eq:FM_AET} are well described by the Fisher matrix formalism.} \label{fig:5dFMcheck} 
\end{figure}

\clearpage
\twocolumngrid
\bibliographystyle{unsrt}
\bibliography{refs}  

\end{document}